\documentclass{JHEP3}
\usepackage{amsmath}
\usepackage{axodraw}
\usepackage{epsfig}
\usepackage{amssymb}
\usepackage{mathrsfs}
\usepackage{mathcomp}
\usepackage{cite}
\usepackage[latin1]{inputenc}
\usepackage{xspace}
\newcommand{\pythia}{P\protect\scalebox{0.8}{YTHIA}\xspace}
\newcommand{\dipsy}{\protect\scalebox{0.8}{DIPSY}\xspace}
\newcommand{\herwig}{\protect\scalebox{0.8}{HERWIG}\xspace}
\newcommand{\pytppp}{P\protect\scalebox{0.8}{YTHIA}8\xspace}
\newcommand{\sherpa}{S\protect\scalebox{0.8}{HERPA}\xspace}
\newcommand{\phobos}{PHOBOS\xspace}
\newcommand{\phenix}{PHENIX\xspace}
\newcommand{\atlas}{ATLAS\xspace}
\newcommand{\alice}{ALICE\xspace}
\newcommand{\cms}{CMS\xspace}
\newcommand{\totem}{TOTEM\xspace}

\providecommand{\eqref}[1]{eq.~(\ref{#1})\xspace}
\newcommand{\eq}[1]{(\ref{#1})\xspace}
\renewcommand{\eqref}[1]{eq.~(\ref{#1})\xspace}
\newcommand{\Eqref}[1]{Eq.~(\ref{#1})\xspace}
\newcommand{\eqsref}[1]{eqs.~(\ref{#1})\xspace}

\newcommand{\fig}[1]{\ref{#1}}
\newcommand{\figref}[1]{figure~\fig{#1}}

\newcommand{\sect}[1]{\ref{#1}}
\newcommand{\sectref}[1]{section~\sect{#1}}
\newcommand{\sectrefs}[1]{sections~\sect{#1}}

\newenvironment{absolutelynopagebreak}
  {\par\nobreak\vfil\penalty0\vfilneg
   \vtop\bgroup}
  {\par\xdef\tpd{\the\prevdepth}\egroup
   \prevdepth=\tpd}

\def\text{\mathrm}
\def\eg{\emph{e.g.}}
\def\ie{\emph{i.e.}}
\def\cf{\emph{c.f.}}

\def\mrm#1{\mathrm{#1}}
\def\mbf#1{\mathbf{#1}}
\def\sub#1{\ensuremath{_{\mrm{#1}}}}
\def\sup#1{\ensuremath{^{\mrm{#1}}}}
\def\subtot{\sub{tot}}
\def\subtoti#1{\ensuremath{_{\mrm{tot},#1}}}

\def\subabs{\sub{abs}}

\def\subineltot{\sub{in}}
\def\subel{\sub{el}}
\def\subqel{\sub{el*}}
\def\subdtot{\sub{D}}
\def\subsdp{\sub{Dp}}
\def\subsdt{\sub{Dt}}
\def\subdd{\sub{DD}}
\def\subdiff{\sub{diff}}
\def\subw{\ensuremath{_w}}
\def\subwinc{\ensuremath{_{w\sub{inc}}}}
\def\subwabs{_{w\sub{abs}}}

\def\winc{\ensuremath{w\sub{inc}}}
\def\wabs{\ensuremath{w\sub{abs}}}
\def\winel{\ensuremath{w\sub{exc}}}
\def\nucpos{\ensuremath{\mbf{b}_\nu}}
\def\nucrelpos{\ensuremath{\mbf{\tilde{b}}_\nu}}

\newcommand{\newdisk}{2$\times2$-disk\xspace}
\newcommand{\newfrit}{FritiofP8\xspace}
\newcommand{\ourabsorptive}{Absorptive\xspace}

\def\llangle{\left\langle}
\def\rrangle{\right\rangle}

\def\pp{\ensuremath{\mrm{pp}}}
\def\pbarp{\ensuremath{\bar{\mrm{p}}\mrm{p}}}
\def\pd{\ensuremath{\mrm{pd}}}
\def\pA{\ensuremath{\mrm{p}A}}

\def\pN{\ensuremath{\mrm{p}N}}
\def\pPb{\ensuremath{\mrm{pPb}}}

\def\AA{\ensuremath{AA}}
\def\NN{\ensuremath{NN}}

\def\Pb{\ensuremath{\mrm{Pb}}}

\def\pd{\ensuremath{\mrm{pd}}}
\def\Npart{\ensuremath{N\sub{part}}}
\def\Ncoll{\ensuremath{N\sub{coll}}}

\def\ColText(#1,#2)[#3]#4{\Text(#1,#2)[#3]{#4}}

\usepackage{ifthen}

\def\showcommentsflag{0}
\newcommand{\showcomments}{\def\showcommentsflag{1}}

\newcounter{commentcounter}%
\newcommand{\comment}[1]{\ifnum\showcommentsflag > 0%
\addtocounter{commentcounter}{1}%
{{\Red{\ensuremath{\ddagger^{\arabic{commentcounter}}}}}}%
\marginpar{\raggedright\tiny\it{{\Red{\ensuremath{\ddagger^{\arabic{commentcounter}}}}} {#1}}}
\fi%
}
\newcommand{\commentdel}[2]{\ifnum\showcommentsflag > 0%
\Red{\sout{#1}}\comment{#2}%
\fi
}
\newcommand{\commentadd}[2]{\ifnum\showcommentsflag > 0%
\comment{#2}\Red{#1}%
\else
#1
\fi
}
\newcommand{\commentchange}[3]{\ifnum\showcommentsflag > 0%
\Red{\sout{#2}}\comment{#3}\Red{#1}%
\else
#1
\fi
}
\newcommand{\nocomment}[1]{\ifnum\showcommentsflag > 0%
{\tiny\it\Red{\{#1}\}}
\fi%
}
\newcommand{\nocommentdel}[1]{\ifnum\showcommentsflag > 0%
\Red{\sout{#1}}%
\fi
}
\newcommand{\nocommentadd}[1]{\ifnum\showcommentsflag > 0%
\Red{#1}%
\else
#1
\fi
}
\newcommand{\nocommentchange}[2]{\ifnum\showcommentsflag > 0%
\Red{\sout{#2}}\Red{#1}%
\else
#1
\fi
}

\showcomments
\keywords{QCD, Nucleus collisions, Fluctuations, Glauber models, Diffraction}
\preprint{LU-TP 16-39\\
  MCnet-16-26\\
  arXiv:1607.04434 [hep-ph]\\
}

\title{Diffractive and non-diffractive wounded nucleons and final
  states in pA collisions\footnote{ Work supported in part by the
    MCnetITN FP7 Marie Curie Initial Training Network, contract
    PITN-GA-2012-315877, and the Swedish Research Council (contracts
    621-2012-2283 and 621-2013-4287)}}

\author{Christian Bierlich, Gösta Gustafson, and Leif Lönnblad \\
  Dept.~of Astronomy and Theoretical Physics,
  Sölvegatan 14A, S-223 62  Lund, Sweden\\
  E-mail: \email{Christian.Bierlich@thep.lu.se},
    \email{Gosta.Gustafson@thep.lu.se},
    and \email{Leif.Lonnblad@thep.lu.se}}
    \abstract{We review the state-of-the-art of Glauber-inspired models
    for estimating the distribution of the number of participating
    nucleons in \pA\ and \AA\ collisions. We argue that there is room
    for improvement in these model when it comes to the treatment of
    diffractive excitation processes, and present a new simple
    Glauber-like model where these processes are better taken into
    account. We also suggest a new way of using the number of
    participating, or wounded, nucleons to extrapolate event characteristics from \pp\ collisions, and hence get an estimate of basic hadronic final-state properties in \pA\ collisions, which may be
    used to extract possible nuclear effects. The new method is inspired by the Fritiof model, but based on the full, semi-hard multiparton interaction model of \pytppp.}

\begin{document}

\sloppy
\newpage
\section{Introduction}

An important topic in the studies of the strong interaction is the understanding of the
features of hot and dense nuclear matter. To correctly interpret
signals for collective behaviour in high energy nucleus--nucleus
collisions, it is necessary to have a realistic extrapolation of the
dynamics in \pp\ collisions. Here experiments on \pA\ collisions have
been regarded as an important intermediate step.  As an example
refs.~\cite{Bzdak:2013zla, McLerran:2015lta} have discussed the
possibility to discriminate between the dynamics of the wounded
nucleon model and that of the Color Glass Condensate formalism in \pPb\
collisions at the LHC.

An extrapolation of results from \pp\ to \pA\ and \AA\ collisions is
generally performed using the Glauber formalism~\cite{Glauber:1955qq,
  Miller:2007ri}. This model is based on the eikonal approximation,
where the interaction is driven by absorption into inelastic channels.
Elastic scattering is then the shadow of absorption, and determined by
the optical theorem. The projectile nucleon(s) are assumed to travel
along straight lines and undergo multiple sub-collisions with nucleons
in the target.
The Glauber model has been commonly used in experiments at RHIC and LHC, \eg\ to
estimate the number of participant nucleons, \Npart, and the number of binary
nucleon--nucleon collisions, \Ncoll, as a function of centrality.
%
A basic assumption is then that one can compare a \pA\ or an \AA\
collision, at a certain centrality with, \eg, \Npart/2 or \Ncoll\
times the corresponding result in \pp\ collisions (for which
\Npart~=~2). A comparison with a fit to \pp\ collision data, folded by
the distribution in \Npart/2 or \Ncoll, can then be used to
investigate nuclear effects on various observables.

There are several problems related to such analyses, and in this paper
we will concentrate on two of them:
\begin{itemize}
\item Since the actual impact parameter is not a physical observable,
  the experiments typically select an observable, which is expected to
  be strongly correlated with the impact parameter (such as a forward
  energy or particle flow).  This implies that the definition of
  centrality becomes detector dependent, which, among other problems,
  also implies difficulties when comparing experimental results with
  each other and with theoretical calculations.
\item When the interaction is driven by absorption, shadow scattering
  (meaning diffraction) can contain elastic \emph{as well as single
    and double diffractive excitation}. This is important since
  experiments at high energy colliders show, that diffractive
  excitation is a significant fraction of the total cross section, and
  not limited to low masses (see \eg\
  \cite{Abe:1993wu,Aad:2012pw,Khachatryan:2015gka}).
%
%
  Thus the driving force in Glauber's formalism should be the
  absorptive, meaning the non-diffractive inelastic cross section, and
  \emph{not the total inelastic} cross section.
\end{itemize}
In the following we will argue that the approximations normally used
in this procedure are much too crude, and we will present a number of
suggestions for how they can be improved, both in the way \Npart\
and \Ncoll\ are calculated and the way \pp\ event characteristics are extrapolated to get reference distributions. 
In both cases we will show that diffractive processes play an important role.

In Glauber's original analysis only elastic scattering was
taken into account, but it was early pointed out by
Gribov~\cite{Gribov:1968jf}, that diffractive excitation of the
intermediate nucleons gives a significant contribution. However,
problems encountered when taking diffractive excitation into account have
implied, that this has frequently been neglected, also in recent applications
(see \eg\ the review by Miller \emph{et al.} \cite{Miller:2007ri}).
Thus the ``black disk'' approximation, and other simplifying treatments,
are still frequently used in analyses of experimental results.\footnote{
The effects of the black disk approximation have also been
discussed in ref.~\cite{Gustafson:2015ada}.}

A way to include diffractive excitation in a Glauber analysis, using the
Good--Walker formalism, was formulated
by Heiselberg \emph{et al.} \cite{Heiselberg:1991is}. It was further developed
in several papers (see refs.~\cite{Blaettel:1993ah, Alvioli:2013vk, 
  Alvioli:2014sba, Alvioli:2014eda} and further references in
there) and is often called the ``Glauber--Gribov'' (GG) model. In the Good--Walker
formalism \cite{Good:1960ba}, diffractive excitation is described as the
result of fluctuations in the nucleon's partonic substructure. When
used in impact parameter space, it has the advantage
that saturation effects can easily be taken into account, which
makes it particularly suited for applications in collisions with nuclei.

The ``Glauber--Gribov'' model has been
applied both to data from RHIC and in recent analyses of data from the LHC,
\eg\ in refs.~\cite{Aad:2015zza, Adam:2014qja}
However, although this formalism implies a significant improvement of the data
analyses, also in this formulation the treatment of diffractive excitation is 
simplified, as the full structure of single excitation of either the projectile 
or the target, and of double diffraction, is not taken into account. As we
will show in this paper, this simplification causes important problems,
and we will here present a very simple model which separates the fluctuations
in the projectile and the target nucleons.

To guide us in our investigation of conventional Glauber models we use the
\dipsy Monte Carlo program \cite{Avsar:2005iz, Avsar:2006jy,
  Flensburg:2011kk}, which is based on Mueller's dipole approach to
BFKL evolution \cite{Mueller:1993rr,Mueller:1994jq}, but also includes
important non-leading effects, saturation and confinement. It
reproduces fairly well both total, elastic, and diffractive \pp\ cross
sections, and has also recently been applied to \pA\
collisions~\cite{Gustafson:2015ada}. The \dipsy model gives a very
detailed picture of correlations and fluctuations in the initial state
of a nucleon, and by combining it with a simple geometrical picture of
the distribution of nucleons in a nucleus in its ground state, we can
build up an equally detailed picture of the initial states in \pA\
and \AA\ collisions. This allows us to gain new insights into the pros
and cons of the approximations made in conventional Glauber Models.

The \dipsy program is also able to produce fully exclusive hadronic
final states in \pp\ collisions, giving a reasonable description of
minimum bias data from \eg\ the LHC \cite{Flensburg:2011kk}. It could,
in principle also be used to directly model final states in \pA\ and
\AA, but due to some shortcomings, we will in this paper instead only
use general features of these final states to motivate a revival of
the old Fritiof model \cite{Andersson:1986gw, Pi:1992ug} with
great similarities with the original "wounded nucleon" model
\cite{Bialas:1976ed}. (For a more recent update of the wounded
nucleon model see ref.~\cite{Bialas:2008zza}.)

For energies up to (and including) those at fixed target experiments
at CERN, the particle density at mid-rapidity in \pp\ collisions is
almost energy independent. For higher energies the density increases,
and the $p_\perp$ distribution gets a tail to larger values.
However, for minimum bias events with lower $p_\perp$, the wounded nucleon
model still works with the multiplicity scaling with the number of
participating (wounded) nucleons, both at RHIC \cite{Back:2004mr,
Adare:2013nff} and LHC \cite{ALICE:2012xs}. For higher $p_\perp$ the
distributions scale, however, better with the number of binary $NN$
collisions, indicating the effect of hard parton-parton sub-collisions
\cite{Adam:2014qja}.

We will here argue that, due to the relatively flat
distribution in rapidity of high-mass diffractive processes, absorbed
and diffractively excited nucleons will contribute to the \pA\ (and in
principle also \AA) final states in very similar ways, as wounded
nucleons.  We will also present preliminary results where we use our
modified GG model to calculate the number distribution of wounded
nucleons in \pA, and from that construct hadronic final states by
stacking diffractive excitation events, on top of a primary
non-diffractive scattering, using \pytppp with its semi-hard
multi-parton interaction picture of hadronic collisions.

Although this remarkably simple picture gives very promising results,
we find that there is a need for differentiating between
diffractively and non-diffractively wounded nucleons. 
We will here be helped by the simple model mentioned above, in which
fluctuations in the projectile and the target nucleon are treated separately.
The model involves treating both the projectile and target
as semi-transparent disks, separately fluctuating between two sizes
according to a given probability. The radii, the transparency and the
fluctuation probability is then adjusted to fit the non-diffractive
nucleon--nucleon cross section, as well as the elastic, single
diffractive and double diffractive cross sections. Even though this is
a rather crude model, it will allow us to investigate effects of the
difference between diffractively and non-diffractively wounded
nucleons.

We will begin this article by establishing in \sectref{sec:framework}
the framework we will use to describe high energy nucleon--nucleon
scattering, with special emphasis on the Good--Walker formalism for
diffractive excitation. In \sectref{sec:glauber} we will then use this
framework to analyse the Glauber formalism in general and define the
concept of a \textit{wounded} target cross section. In
\sectref{sec:models} we dissect the conventional Glauber models and the
Glauber--Gribov model together with the \dipsy model and present some
comparisons of the resulting number distributions of wounded nucleons
in \pA. In \sectref{sec:model-final-stat} we then go on to present our
proposed model for constructing fully exclusive hadronic final states, and
compare the procedure to recent results on particle distributions in
\pA\ collisions from the LHC, before we present conclusions and an
outlook in \sectref{sec:outlook}.

\section{Dynamics of high energy \pp\ scattering}
\label{sec:framework}

\subsection{Multiple sub-collisions and perturbative parton--parton interaction}

As mentioned in the introduction, at energies up to those at fixed
target experiments and the ISR at CERN, the \pp\ cross sections and
particle density, $dn/dy$ are relatively independent of energy. For
collisions with nuclei the wounded nucleon model works quite well
\cite{Bialas:1976ed}, which formed the basis for the development of
the Fritiof model~\cite{Andersson:1986gw}. This model worked very
well within that energy range, but at higher energies it could not
in a satisfactory way reproduce the development of a high $p_\perp$ tail
caused by hard parton-parton interactions. Nevertheless the wounded
nucleon model works well for minimum bias events even at
LHC energies, if the rising rapidity plateau in \pp\ collisions is
taken into account, although the production of high $p_\perp$ particles
appear to scale better with the number of $NN$ collisions.
These features may be interpreted as signals for dominance of soft
interactions, and were the basis for the development of the Fritiof
model~\cite{Andersson:1986gw}. This model worked very well within that
energy range, but at higher energies, available at \pbarp\
colliders at CERN and Fermilab, the effects of (multiple) hard
parton--parton sub-collisions became increasingly important, and not so
easily incorporated in the Fritiof model.

Today high energy collisions (above $\sqrt{s} \sim$ 100~GeV) are more
often described as the result of multiple partonic sub-collisions,
described by perturbative QCD. This picture was early proposed by
Sj\"ostrand and van Zijl \cite{Sjostrand:1987su}, and is implemented
in the \pytppp event generator \cite{Sjostrand:2014zea}.
This picture has also
been applied in other generators such as \herwig \cite{Bahr:2008pv},
\sherpa \cite{Gleisberg:2008ta}, \dipsy
\cite{Avsar:2005iz,Flensburg:2011kk}, and others. The dominance of
perturbative effects can here be understood from the suppression of
low-$p_\perp$ partons due to saturation, as expressed \eg\ in the Color
Glass Condensate formalism \cite{McLerran:1993ni}.

\subsection{Saturation and the transverse coordinate space}

\subsubsection{The eikonal approximation}

The large cross sections in hadronic collisions imply that unitarity
constraints are important, and the elastic amplitude has to satisfy
the optical theorem, which with convenient normalisation reads
\begin{equation}
\mathrm{Im}\, A\subel = \frac{1}{2} \left\{ | A\subel|^2 + \sum_j|A_j|^2 \right\}.
\label{eq:optical}
\end{equation}
Here the sum runs over all inelastic channels $j$. In high energy \pp\
collisions the real part of the elastic amplitude is small, which indicates
that the interaction is dominated by absorption into inelastic channels, with
elastic scattering formed as the diffractive shadow of this absorption.
This diffractive scattering is dominated by small $p_\perp$, and the scattered
proton continues essentially along its initial direction.

At high energies and small transverse momenta, multiple scattering corresponds
to a convolution in transverse momentum 
space, which is represented by a product in transverse coordinate space. This
implies that diffraction and rescattering is more easily described in impact
parameter space. 
In a situation where all inelastic channels correspond to absorption (meaning
\emph{no diffractive excitation}),
the optical theorem in 
\eqref{eq:optical} implies that the elastic amplitude in impact
parameter space is given by
\begin{equation}
A\subel(b)= i \left\{1 -\sqrt{1- P\subabs(b)}\right\}.
\label{eq:Ael}
\end{equation}
Here $P\subabs(b)=\sum_j|A_j(b)|^2$ represents the probability for absorption
into inelastic channels. 

If the absorption probability in the 
Born approximation is given by $2F(b)$, then unitarity is restored by 
rescattering effects, which exponentiates in $\mathbf{b}$-space and give the
eikonal approximation:
\begin{equation}
P\subabs = d \sigma\subabs/d^2b = 1-e^{-2F(b)},
\label{eq:sigmaabs}
\end{equation}
To simplify the
notation we introduce the nearly real amplitude $T= -iA\subel = 1-S$.
The relation in \eqref{eq:Ael} then gives $S(b)=e^{-F(b)}$ and $T(b)=
1-e^{-F(b)}$. The optical theorem then gives
\begin{eqnarray}
T = 1 - &S& = 1-e^{-F} \nonumber\\
d\sigma\subel/d^2b &  =& T^2=(1-e^{-F})^2 \nonumber\\
d\sigma\subtot/d^2b&  =& 2T=2(1-e^{-F}).
\label{eq:sigmael-tot}
\end{eqnarray} 
We note that the possibility of diffractive excitation is not included
here. Therefore the absorptive cross section in \eqref{eq:sigmaabs} is
the same as the inelastic cross section.

How to include diffractive excitation and its relation to fluctuations
will be discussed below in \sectref{sec:goodwalker}. We then also note
that diffractive excitation is very sensitive to saturation effects,
as the fluctuations go to zero when saturation drives the interaction
towards the black limit.

That rescattering exponentiates in transverse coordinate space also
makes this formulation suitable for generalisations to collisions with
nuclei.

\subsubsection{Dipole models in transverse coordinate space}

In this paper we will use our implementation of Mueller's dipole
model, called \dipsy, in order to have a model which gives a realistic
picture of correlations and fluctuations in the colliding nucleons. In
this way we can evaluate to what extent Glauber-like models are able
to take such effects into account. The \dipsy model has been described
in a series of papers
\cite{Avsar:2005iz,Avsar:2006jy,Flensburg:2011kk} and we will here
only give a very brief description.  Mueller's dipole model
\cite{Mueller:1993rr,Mueller:1994jq} is a formulation of LL BFKL
evolution in impact parameter space. A colour charge is always
screened by an accompanying anti-charge. A charge--anti-charge pair
can emit bremsstrahlung gluons in the same way as an electric dipole,
with a probability per unit rapidity for a dipole ($\mathbf{r}_0,
\mathbf{r}_1)$ to emit a gluon in the point $\mathbf{r}_2$, given by
(\cf\ \figref{fig:dipolesplit})
\begin{equation}
	\label{eq:cascade}
\frac{d\mathcal{P}}{dy} = \frac{\bar{\alpha}}{2\pi} d^2 \mathbf{r}_2
      \frac{r_{01}^2}{r_{02}^2 \,r_{12}^2}.
\end{equation}
The important difference from electro-magnetism is that the emitted
gluon carries away colour, which implies that the dipole splits in two
dipoles. These dipoles can then emit further gluons in a cascade,
producing a chain of dipoles as illustrated in
\figref{fig:dipolesplit}.
 
\FIGURE{
\centering
 \scalebox{0.68}{\mbox{
        \begin{picture}(400,100)(0,0)
          \Vertex(50,100){2}
          \Vertex(50,0){2}
          \Vertex(150,100){2}
          \Vertex(150,0){2}
          \Vertex(180,30){2}
          \Vertex(285,100){2}
          \Vertex(285,0){2}
          \Vertex(315,30){2}
          \Vertex(315,90){2}
          \Line(50,0)(50,100)
          \DashLine(150,100)(150,0){2}
          \Line(150,100)(180,30)
          \Line(150,0)(180,30)
          \DashLine(285,100)(285,0){2}
          \DashLine(285,100)(315,30){2}
          \Line(285,100)(315,90)
          \Line(315,90)(315,30)
          \Line(285,0)(315,30)
          \Text(40,100)[]{$Q$}
          \Text(40,0)[]{$\bar{Q}$}
          \Text(60,100)[]{$1$}
          \Text(60,0)[]{$0$}
          \Text(140,100)[]{$1$}
          \Text(140,0)[]{$0$}
          \Text(140,50)[]{$\mathbf{r}_{01}$}
          \Text(190,30)[]{$2$}
          \Text(175,78)[]{$\mathbf{r}_{12}$}
          \Text(175,2)[]{$\mathbf{r}_{02}$}
        \Text(275,100)[]{$1$}
          \Text(275,0)[]{$0$}
          \Text(325,30)[]{$2$}
          \Text(325,90)[]{$3$}
          \LongArrow(75,50)(110,50)
          \LongArrow(210,50)(245,50)
          \LongArrow(360,25)(360,50)
          \LongArrow(360,25)(385,25)
          \Text(360,60)[]{$y$}
          \Text(395,25)[]{$x$}
        \end{picture}
      }}
    \caption{\label{fig:dipolesplit}A colour dipole cascade in
      transverse coordinate space. A dipole can radiate a gluon. The
      gluon carries away colour, which implies that the dipole is
      split in two dipoles, which in the large $N_c$ limit radiate
      further gluons independently.}
}

When two such chains, accelerated in opposite directions, meet, they
can interact via gluon exchange. This implies exchange of colour, and
thus a reconnection of the chains as shown in
\figref{fig:dipolescattering}.

\FIGURE{ \centering
  \scalebox{1}{\mbox{
\begin{picture}(300,80)(-50,0)

\Text(27,35)[r]{{\footnotesize $i$}}
\Text(53,37)[l]{{\footnotesize $j$}}
\Text(30,29)[tl]{{\footnotesize 2}}
\Text(30,41)[bl]{{\footnotesize 1}}
\Text(51,44)[br]{{\footnotesize 3}}
\Text(50,31)[tr]{{\footnotesize 4}}
\Vertex(150,32){1}
\Vertex(130,30){1}
\Vertex(130,40){1}
\Vertex(151,43){1}
\Vertex(50,32){1}
\Vertex(30,30){1}
\Vertex(30,40){1}
\Vertex(51,43){1}

\LongArrowArcn(90,20)(20,120,60)

\Line(10,20)(30,30)
\ArrowLine(30,40)(30,30)
\Line(30,40)(20,50)
\Line(20,50)(28,57)
\Line(28,57)(18,70)

\Line(70,20)(60,10)
\Line(60,10)(50,32)
\ArrowLine(50,32)(51,43)
\Line(51,43)(60,52)
\Line(60,52)(60,65)

\ArrowLine(130,30)(110,20)
\ArrowLine(120,50)(130,40)
\Line(120,50)(128,57)
\Line(128,57)(118,70)

\Line(170,20)(160,10)
\ArrowLine(160,10)(150,32)
\ArrowLine(151,43)(160,52)
\Line(160,52)(160,65)

\ArrowLine(150,32)(130,30)
\ArrowLine(130,40)(151,43)

\end{picture}
}}
\caption{\label{fig:dipolescattering}In a collision between two dipole
  cascades, two dipoles can interact 
  via gluon exchange. As the exchanged gluon carries colour, the two dipole
  chains become recoupled.} 
}

The elastic scattering amplitude for gluon exchange is in the Born approximation
given by
\begin{equation}
	\label{eq:interaction}
f_{ij}=\frac{\alpha_s^2}{2} \ln^2\left(\frac{r_{13} r_{24}}{r_{14} r_{23}}\right).
\end{equation}
BFKL evolution is a stochastic process, and many sub-collisions may occur 
independently. Summing over all possible pairs gives the total Born amplitude
\begin{equation}
F=\sum_{ij} f_{ij}.
\end{equation}
The unitarised amplitude then becomes
\begin{equation}
T=1-e^{-\sum f_{ij}},
\end{equation}
and the cross sections are given by
\begin{equation}
	\label{eq:xsec}
d\sigma\subel/d^2b=T^2,\,\,\,\,\,\,\,\,d\sigma\subtot/d^2b=2T
\end{equation}

\subsubsection{The Lund dipole model \dipsy}

The \dipsy model \cite{Avsar:2005iz,Avsar:2006jy,Flensburg:2011kk} is
a generalisation of Mueller's cascade, which includes a set of
corrections:
\begin{itemize}
\item Important non-leading effects in BFKL evolution.\\
  Most essential are those related to energy conservation and running
  $\alpha_s$.

\item Saturation from Pomeron loops in the evolution.\\
  Dipoles with identical colours form colour quadrupoles, which give
  Pomeron loops in the evolution. These are not included in Mueller's
  model or in the BK equation.

\item Confinement via a gluon mass satisfies $t$-channel unitarity.

\item It can be applied to collisions between electrons, protons, and nuclei.
\end{itemize}

Some results for \pp\ total and elastic cross sections are shown in
refs.\ \cite{Flensburg:2008ag,Flensburg:2010kq}.  We note that there
is no input structure functions in the model; the gluon distributions
are generated within the model. We also note that the elastic cross
section goes to zero in the dip of the $t$-distribution, as the real
part of the amplitude is neglected.

\subsection{Diffractive excitation and the Good--Walker formalism}
\label{sec:goodwalker}

In his analysis of the Glauber formalism, Gribov considered low mass
excitation in the resonance region, but experiments at high energy
colliders have shown, that diffractive excitation is not limited to
low masses, and that high mass diffraction is a significant fraction
of the \pp\ cross section also at high energies (see \eg\
\cite{Abe:1993wu,Aad:2012pw,Khachatryan:2015gka}).  Diffractive
excitation is often described within the Mueller--Regge formalism
\cite{Mueller:1970fa}, where high-mass diffraction is given by a
triple-Pomeron diagram. Saturation effects imply, however, that
complicated diagrams with Pomeron loops have to be included, which
leads to complicated resummation schemes, see \eg\
refs.~\cite{Gotsman:2014pwa, Khoze:2014aca, Ostapchenko:2010vb}. These
effects make the application in Glauber calculations quite difficult.

High mass diffraction can also be described, within the Good--Walker
formalism \cite{Good:1960ba}, as the result of fluctuations in the
nucleon's partonic substructure. Diffractive excitation is here
obtained when the projectile is a linear combination of states with
different absorption probabilities.  This formalism was first applied
to \pp\ collisions by Miettinen and Pumplin \cite{Miettinen:1978jb},
and later within the formalism for QCD cascades by Hatta \emph{et al.}
\cite{Hatta:2006hs} and by Avsar and coworkers \cite{Avsar:2007xg,
  Flensburg:2010kq}. When used in impact parameter space, this
formulation has the advantage that saturation effects can easily be
taken into account, and this feature makes it particularly suited in
applications for collisions with nuclei.  (For a BFKL Pomeron, the
Good--Walker and the Mueller--Regge formalisms describe the same
physics, seen from different sides \cite{Gustafson:2012hg}.)

As an illustration of the Good--Walker mechanism, we can study a
photon in an optically active medium. For a photon beam passing a
black absorber, the waves around the absorber are scattered
elastically, within a narrow forward cone.  In the optically active
medium, right-handed and left-handed photons move with different
velocities, meaning that they propagate as particles with different
mass. Study a beam of right-handed photons hitting a polarised target,
which absorbs photons polarised in the $x$-direction. The
diffractively scattered beam is then a mixture of right- and
left-handed photons. If the right-handed photons have lower mass, this
means that the diffractive beam contains also photons excited to a
state with higher mass.

\subsubsection{A projectile with substructure colliding with a  structureless target}

For a projectile with a substructure, the mass eigenstates can differ
from the eigenstates of diffraction. Call the diffractive eigenstates
$\Phi_k$, with elastic scattering amplitudes $T_k$. The mass
eigenstates $\Psi_{i}$ are linear combinations of the states $\Phi_k$:
\begin{equation}
\Psi_{i} = \sum_k  c_{ik} \Phi_k\,\,\,\,\,\,(\mathrm{with}\,\,\Psi_{in}=\Psi_1).
\label{eq:eigenstates}
\end{equation}
The elastic scattering amplitude is given by
\begin{equation}
\llangle \Psi_{1} | T | \Psi_{1} \rrangle = \sum c_{1k}^2 T_k 
= \llangle T \rrangle,
\end{equation}
and the elastic cross section
\begin{equation}
d \sigma\subel/d^2 b = \left(\sum c_{1k}^2 T_k\right)^2 = \llangle T\rrangle ^2.
\label{eq:sigmael}
\end{equation}
The amplitude for diffractive transition to the mass eigenstate $\Psi_k$ is
given by 
\begin{equation}
\llangle \Psi_{i} | T | \Psi_{1} \rrangle = \sum_k  c_{ik} T_k c_{1k},
\end{equation}
which gives a total diffractive cross section (including elastic
scattering)
\begin{equation}
d\sigma\subdiff/d^2 b=\sum_i \llangle \Psi_{1} | T | \Psi_{i} \rrangle \llangle
\Psi_{i} | T | \Psi_{1} \rrangle =\llangle T^2 \rrangle.
\end{equation}
Consequently the cross section for diffractive excitation is
\emph{given by the fluctuations}:
\begin{equation}
d\sigma\subdtot/d^2 b  = d\sigma\subdiff- d \sigma\subel =
\llangle T^2 \rrangle - \llangle T \rrangle ^2.
\end{equation}
We note in particular that in this case the \emph{absorptive cross
  section equals the inelastic non-diffractive cross
  section}. Averaging over different eigenstates \eqref{eq:sigmaabs}
gives
\begin{eqnarray}
d \sigma\subabs/d^2b &=& \llangle 1-e^{-2F(b)}\rrangle = 
\llangle 1-(1-T)^2 \rrangle = 2 \llangle T \rrangle - \llangle T^2 \rrangle 
\nonumber\\
&=& d\sigma\subtot/d^2b - d\sigma\subdiff/d^2 b.
\label{eq:sigmaabsGW}
\end{eqnarray}

\subsubsection{A target with a substructure}
\label{sec:substructure}
If also the target has a substructure, it is possible to have either
single excitation of the projectile, of the target, or double
diffractive excitation. Let $\Psi_k^{(p)}$ and $\Psi_l^{(t)}$ be the
diffractive eigenstates for the projectile and the target
respectively, and $T_{kl}$ the corresponding eigenvalue. (We here make
the assumption that the set of eigenstates for the projectile are the
same, for all possible target states. This assumption is also made in
the \dipsy model discussed above.) The total diffractive cross
section, including elastic scattering, is then obtained by taking the
average of $T_{kl}^2$ over all possible states for the projectile and
the target. Subtracting the elastic scattering then gives the total
cross section for diffractive excitation:
\begin{equation}
d\sigma\subdtot/d^2 b   = \llangle T^2 \rrangle_{p,t} - (\llangle T
\rrangle_{p,t} )^2. 
\label{eq:totaldiff}
\end{equation}
Here the subscripts $p$ and $t$ denote averages over the projectile
and target respectively.

Taking the average over target states before squaring gives the
probability for an elastic interaction for the target. Subtracting
single diffraction of the projectile and the target from the total in
\eqref{eq:totaldiff} will finally give the double diffraction. Thus we
get the following relations:
\begin{eqnarray}
d\sigma\subtot/d^2 b& =&2\,\llangle T \rrangle_{p,t}\nonumber\\
d\sigma\subel/d^2 b& =&\llangle T \rrangle_{p,t}^2\nonumber\\
 d\sigma\subsdp/d^2 b &=&\llangle\llangle T \rrangle_{t}^2\rrangle_p-
  \llangle T \rrangle_{p,t}^2\nonumber\\
 d\sigma\subsdt/d^2 b &=&\llangle\llangle T \rrangle_{p}^2\rrangle_t-
  \llangle T \rrangle_{p,t}^2\nonumber\\
 d\sigma\subdd/d^2 b& =& \llangle T^2 \rrangle_{p,t}-\llangle\llangle T
 \rrangle_{t}^2\rrangle_p - \llangle\llangle T \rrangle_{p}^2\rrangle_t +
\llangle T \rrangle_{p,t}^2,
\label{eq:sigmadiff}
\end{eqnarray}
where $\sigma\subsdp$ and $\sigma\subsdt$ is single diffractive
excitation of the projectile and target respectively and
$\sigma\subdd$ is double diffractive excitation.  Also here the
absorptive cross section, which will be important in the following
discussion of the Glauber model, corresponds to the
\emph{non-diffractive} inelastic cross section:
\begin{equation}
d\sigma\subabs/d^2 b =2 \llangle T \rrangle_{p,t} - \llangle T ^2\rrangle_{p,t}.
\label{eq:sigmadiffabs}
\end{equation}

\subsubsection{Diffractive eigenstates at high energies}

In the early work by Miettinen and Pumplin \cite{Miettinen:1978jb},
the authors suggested that the diffractive eigenstates correspond to
different geometrical configurations of the valence quarks, as a
result of their relative motion within a hadron.  At higher energies
the proton's partonic structure is dominated by gluons.  The BFKL
evolution is a stochastic process, and it is then natural to interpret
the perturbative parton cascades as the diffractive eigenstates (which
may also depend on the positions of the emitting valence
partons). This was the assumption in the work by Hatta \emph{et al.}
\cite{Hatta:2006hs} and in the \dipsy model. Within the \dipsy model,
based on BFKL dynamics, it was possible to obtain a fair description
of both the experimental cross section \cite{Avsar:2007xg,
  Flensburg:2010kq} and final state properties \cite{Flensburg:2012zy}
for diffractive excitation.  In the GG model two sources to
fluctuations are considered; first fluctuations in the geometric
distribution of valence quarks, and secondly fluctuations in the
emitted gluon cascades, called colour fluctuations or flickering. In
ref.~\cite{Alvioli:2014sba} it was concluded that the latter is
expected to dominate at high energies.

We here also note that at very high energies, when saturation drives
the interaction towards the black limit, the fluctuations go to
zero. This implies that diffractive excitation is largest in
peripheral collisions, where saturation is less effective. This is
true both for \pp\ collisions and collisions with nuclei. (Although
diffractive excitation of the projectile is almost zero in central
\pA\ collisions, this is not the case for nucleons in the target.)
\section{Glauber formalism for collisions with nuclei}
\label{sec:glauber}
\subsection{General formalism}
\label{sec:general}

High energy nuclear collisions are usually analysed within the Glauber
formalism~\cite{Glauber:1955qq} (for a more recent overview see
\cite{Miller:2007ri}). In this formalism, target nucleons are treated
as independent, and any interaction between them is
neglected\footnote{In the \dipsy model gluons with the same colour can
  interfere, also when they come from different nucleons.  This
  so-called inter-nucleon swing mechanism was shown
  \cite{Gustafson:2015ada} to have noticeable effects in
  photon--nucleus collisions, but in \pA, especially for heavy nuclei,
  the effects were less that 5\%. We have therefore chosen to ignore
  such effects in this paper, but may return to the issue in a future
  publication.  }. The projectile nucleon(s) travel along straight
lines, and undergo multiple diffractive sub-collisions with small
transverse momenta.  As mentioned in the introduction, multiple
scattering, which in transverse momentum space corresponds to a
convolution of the scattering $S$-matrices, corresponds to a product
in transverse coordinate space.  Thus the matrices $S^{(\pN_\nu)}$,
for the encounters of the proton with the different nucleons in the
target nucleus, factorise:
\begin{equation}
S^{(\pA)} = 
\prod_{\nu=1}^A  S^{(\pN_\nu)}.
\label{eq:targfactorize}
\end{equation}

We denote the impact parameters for the projectile and for the
different nucleons in the target nucleus by $\mbf{b}$ and $\nucpos$
respectively, and define $\nucrelpos \equiv \mbf{b} - \nucpos$.  Using
the notation in \eqref{eq:sigmael-tot}, we then get the following
elastic scattering amplitude for a proton hitting a nucleus with $A$
nucleons:
\begin{equation}
T^{(\pA)}(\mbf{b}) = 1 - \prod_{\nu=1}^A
S^{(\pN_\nu)}(\nucrelpos) =1 - \prod_\nu
\left(1-T^{(\pN_\nu)}(\nucrelpos)\right) = 1 - e^{-\sum_\nu
  F^{(\pN_\nu)}(\nucrelpos)}. 
\label{eq:glauber}
\end{equation}
If there are no fluctuations, neither in the \pp\ interaction nor in
the distribution of nucleons in the nucleus, a knowledge of the
positions $\nucpos$ and the \pp\ elastic amplitude
$T^{(\pp)}(\mbf{\tilde{b}})$ would give the total and elastic \pA\
cross sections via the relations in \eqref{eq:sigmael-tot}:
\begin{eqnarray}
\sigma\subtot^{(\pA)}&=&2\int d^2 b\,T^{(\pA)}(\mbf{b})\\
\sigma\subel^{(\pA)}&=&\int d^2 b\,\left(T^{(\pA)}(\mbf{b})\right)^2
\end{eqnarray}
The inelastic cross section (now equal to the absorptive) would be equal to
the difference 
between these two, in accordance with \eqref{eq:sigmaabs}. 

Fluctuations in the \pp\ interaction are discussed in the following
subsection.  Fluctuations and correlations in the nucleon distribution
within the nucleus are difficult to treat analytically, and therefore
most easily studied by means of a Monte Carlo, as discussed further in
\sectrefs{sec:nucleusgeometry}, \sect{sec:models} and
\sect{sec:model-final-stat} below. Valuable physical insight can,
however, be gained in an approximation where all correlations between
target nucleons are neglected. Such an approximation, called the
optical limit, is discussed in \sectref{sec:uncorrelated}.

\subsection{Gribov corrections. Fluctuations in the \pp\ interaction}
\label{sec:gribov}

Gribov pointed out that the original Glauber model gets significant
corrections due to possible diffractive excitation. In the literature
it is, however, common to take only diffractive excitation of the
projectile into account, disregarding possible excitation of the
target nucleons. In this section we will develop the formalism to
account for excitations of nucleons in both projectile and target. We
will then see that in many cases fluctuations in the target nucleons
will average out, while in other cases they may give important
effects.  (Fluctuations in both projectile and target will, however,
be even more essential in nucleus--nucleus collisions, which we plan
to discuss in a future publication.)

\subsubsection{Total and elastic cross sections}

When the nucleons can be in different diffractive eigenstates, the
amplitudes $T^{(\pN_\nu)}$ in \eqref{eq:glauber} are matrices
$T^{(\pN_\nu)}_{k,l_\nu}$, depending on the states $k$ for the
projectile and $l_\nu$ for the target nucleon $\nu$. The elastic \pA\
amplitude, $\llangle T^{(\pA)}(\mbf{b})\rrangle$, can then still be
calculated from \eqref{eq:glauber}, by averaging over all values for
$k$ and $l_\nu$, with $\nu$ running from 1 to $A$.  Thus
\begin{eqnarray}
d\sigma\subtot^{(\pA)}/d^2b&=&2\,\llangle  T^{(\pA)}(\mbf{b})\rrangle= 2\left\{1-\llangle
S^{(\pA)}(\mbf{b})\rrangle \right\},\label{eq:totgribov}\\
d\sigma\subel^{(\pA)}/d^2b&=&\llangle  T^{(\pA)}(\mbf{b})\rrangle^2. 
\label{eq:elgribov}
\end{eqnarray}
When evaluating the averages in these equations, it is essential that
the projectile proton stays in the same diffractive eigenstate,
$\Phi_k$, throughout the whole passage through the target nucleus,
while the states, $\Phi_{l_\nu}$, for the nucleons in the target
nucleus are uncorrelated from each other.  This implies that for a
\emph{fixed} projectile state $k$, the average of the $S$-matrix over
different states, $l_\nu$, for the target nucleons factorise in
\eqref{eq:targfactorize} or \eq{eq:totgribov}. Thus we have
\begin{equation}
\llangle\, S^{(\pA)}(\mbf{b}) \rrangle =
\llangle\,\llangle \prod_\nu  \llangle
S^{(\pp,\nu)}_{k,l_\nu}(
\nucrelpos)\rrangle_{l_\nu}\rrangle_{\nucpos} \rrangle_{k}. 
\label{eq:targaverage}
\end{equation}
Here $\llangle\cdots\rrangle_k$ ($\llangle\cdots\rrangle_{l_\nu}$)
denotes average over projectile (target nucleon) substructures $k$
($l_\nu$), while $\llangle\cdots\rrangle_{\nucpos}$ denotes average
over the target nucleon positions $\nucpos$, an as before
$\nucrelpos\equiv \mbf{b} - \nucpos$.  We introduce the following
notation for the average of the \pp\ amplitude over target states:
\begin{equation}
T^{(\pp)}_{k}(\nucrelpos)\equiv
\llangle T^{(\pp)}_{k,l}(\nucrelpos)\,\rrangle_l=
\llangle (1- S^{(\pp)}_{k,l}(\nucrelpos))\,\rrangle_l.
\label{eq:averl}
\end{equation}
The \pA\ amplitude can then be written in the form
\begin{equation}
\llangle T_k^{(\pA)}(\mbf{b})\rrangle_k =
\llangle \,\left\{1-\prod_\nu S^{(\pp)}_{k}(\nucrelpos)\right\}\,
\rrangle_{\!\nucpos,k}
=\llangle \,\left\{1-\prod_\nu \left(1-
    T^{(\pp)}_{k}(\nucrelpos)\right)\right\}\, 
\rrangle_{\!\nucpos,k},
\label{eq:targaverage2}
\end{equation}
where the average is taken over the target nucleon positions $\nucpos$
and the projectile states, $k$. The total and elastic cross sections
in \eqsref{eq:totgribov} and \eq{eq:elgribov} are finally obtained
from \eqref{eq:sigmael-tot}.  We want here to emphasise that these
expressions only contain the first moment with respect to the
fluctuations in the target states, $l_\nu$, but also all higher
moments of the fluctuations in the projectile states, $k$.

To evaluate the $b$-integrated cross sections, we must know both the
distribution of the (correlated) nucleon positions, $\nucpos$, and the
$b$-dependence of the \pp\ amplitude $T^{(\pp)}_{k}(b)$. The
distribution of nucleon positions is normally handled by a Monte
Carlo, as will be discussed in \sectref{sec:nucleusgeometry}.  When
fluctuations and diffractive excitation was neglected in
\sectref{sec:general}, the $b$-dependence of $T^{(\pp)}(b)$ could be
well approximated by a Gaussian distribution $C\exp{(-b^2/2B)}$,
corresponding to an exponential elastic cross section $d\sigma/dt
\propto \exp{(Bt)}$. With fluctuations it is necessary to take the
unitarity constraint $T\le 1$ into account, which implies that a large
cross section must be associated with a wider distribution. One should
then check that after averaging the differential elastic cross section
reproduces the observed slope.\footnote{In ref.~\cite{Alvioli:2013vk}
  unitarity is satisfied assuming the slope $B$ to be proportional to
  the fluctuating total cross section $\sigma\subtot$.}
\subsection{Interacting nucleons}
\label{sec:wounded}

\subsubsection{Specification of "wounded" nucleons}

The notion of ``wounded'' nucleons was introduced by Bia\l as,
Bleszy\'{n}ski, and Czy\.{z} in 1976 \cite{Bialas:1976ed}, based on
the idea that inelastic \pA\ or \AA\ collisions can be described as a
sum of independent contributions from the different participating
nucleons\footnote{This idea was also the basis for the Fritiof model
  \cite{Andersson:1986gw}, which has been quite successful for low
  energies.}. In ref.~\cite{Bialas:1976ed} diffractive excitation was
neglected, and thus ``wounded nucleons'' was identical to
inelastically interacting nucleons\footnote{It was also pointed out
  that for \pA\ collisions the number of participant nucleons, $w$,
  and the number of \NN\ sub-collisions, $v$, are related, $v=w+1$,
  and a relation between particle multiplicity and the number of
  wounded nucleons, $w$, is equivalent to a relation to the number of
  \NN\ sub-collisions, $v=w+1$. Only in \AA\ collisions is it possible
  to distinguish a dependence on the number of participating nucleons
  from a dependence on the number of nucleon--nucleon
  sub-collisions.}.

Although the importance of diffractive excitation was pointed out by
Gribov already in 1968~\cite{Gribov:1968jf}, it has, as far as we
know, never been discussed whether or not diffractively excited
nucleons should be regarded as wounded. These nucleons contribute to
the inelastic, but not to the absorptive cross section, as defined in
\eqref{eq:sigmadiffabs}.

Diffractive excitation is usually fitted to a distribution
proportional to $d M_X^2/(M_X^2)^{1+\epsilon}$. A bare triple-Pomeron
diagram would give $\epsilon=\alpha_\mathbb{P}(0)-1$, where
$\alpha_\mathbb{P}(0)$ is the intercept of the Pomeron trajectory,
estimated to around 1.2 from the HERA structure functions at small
$x$. More complicated diagrams tend, however, to reduce
$\epsilon$. (In ref.~\cite{Ostapchenko:2010vb} it is shown that the
largest correction is a four-Pomeron diagram, which gives a
contribution with $\epsilon=0$.)  Fits to LHC data
\cite{Aad:2012pw,Khachatryan:2015gka} give $\epsilon \approx 0.1$, but
with rather large uncertainties.

If $\epsilon$ is small, diffractively excited target nucleons can
contribute to particle production both in the forward and in the
central region. If $\epsilon$ instead is large, diffraction would
contribute mainly close to the nucleus fragmentation region. For
$\epsilon \approx 0.1$, the experimentally favoured value, the
contribution in the central region would be suppressed by a factor
$\exp(-0.1\cdot \Delta \eta) \sim 1/2$ for \pPb\ collisions at LHC.
We conclude that the definition of wounded nucleons should depend
critically upon both the experimental observable studied in a certain
analyses, and upon the still uncertain $M_X$-dependence of diffractive
excitation at LHC energies.  (In \sectref{sec:finalstates} we will
show that a simple model, assuming similar contributions from absorbed
and diffractively excited nucleons actually quite successfully
describes the final state in \pPb\ collisions at LHC.)

Below we present first results for
the absorbed, non-diffractive, nucleons, followed by results when
diffractively excited nucleons are included. 

\subsubsection{Wounded nucleon cross sections}
\label{sec:woundedcross}

\textbf{Absorptive cross section}
\vspace{2mm}

We first assume that wounded nucleons correspond to nucleons absorbed
via gluon exchange, which for large values of $\epsilon$ would be
relevant for observables in the central region, away from the nucleus
fragmentation region.  Due to the relation $T=1-S$, the absorptive
cross section in \eqref{eq:sigmadiffabs} can also be written
$d\sigma\subabs/d^2b= \llangle 1 - S^2 \rrangle$.  We here note that,
as the $S$-matrix factorises in the elastic amplitude in
\eqsref{eq:targfactorize} and \eq{eq:totgribov}, this is also the case
for $S^2$. This implies that
\begin{equation}
\left(S^{(\pA)}_{k,\{l_\nu\}}\right)^2 =
\prod_{\nu=1}^{A}\left(S^{(\pN_\nu)}_{k,l_\nu}\right)^2. 
\label{eq:absproduct}
\end{equation} 
In analogy with \eqref{eq:averl} for $\sigma\subtot$, also here, when
taking the average over the target states $l_\nu$, the factors in the
product depend only on the projectile state $k$ and the positions
$\nucrelpos$. We here introduce the notation
\begin{equation}
W^{(\wabs)}_k(\nucrelpos)\equiv
\llangle \,1-\left(\,S^{(\pp)}_{k,l}(\nucrelpos)\,\right)^2 \,\rrangle_l.
\label{eq:averl2}
\end{equation}
This quantity represents the \emph{probability that nucleon $\nu$ is
  absorbed by a projectile in state $k$}.  Averaging over all values
for $k$ and $\nucpos$, it gives the total \pA\ absorptive, meaning
inelastic non-diffractive, cross section
\begin{equation}
d\sigma^{\pA}\subabs(\mbf{b})/d^2b=\llangle\, \llangle\,\left\{1-
  \prod_\nu\left(1- W^{(\wabs)}_k(\nucrelpos)\right)\right\}\,
\rrangle_{\nucpos}\,\rrangle_k. 
\label{eq:sigmaabspA}
\end{equation}
This expression equals the probability that at least one target
nucleon is absorbed.  

\vspace{2mm}

\textbf{Cross section including diffractively excited target nucleons}
\vspace{2mm}

We now discuss the situation when also diffractively excited target nucleons
should be counted as wounded. (The case with an
excited projectile proton is discussed below.) The probability for
a nucleon, $\nu$, in the nucleus to be diffractively excited
is obtained from \eqref{eq:sigmadiff} by adding single and double
diffraction:  
\begin{eqnarray}
P\sub{D,\nu}&=&
\llangle\left( T^{(\pp)}(\nucrelpos )\right)^2\rrangle_{k,l_\nu} - 
\llangle\left(\llangle \,T^{(\pp)}(\nucrelpos )\,
\rrangle_{l_\nu}\right)^2\rrangle_k\nonumber\\
&=&\llangle\,\llangle\, S^2\rrangle_{l_\nu} -\llangle S
\rrangle_{l_\nu}^2\,\rrangle_k.  
\end{eqnarray}
Adding the absorptive cross section in \eqref{eq:sigmadiffabs} we
obtain the total probability that a target nucleon, $\nu$, is excited
or broken up by either diffraction or absorption,
\begin{equation}
P_{\winc,\nu} =
1- \llangle\, \llangle S\rrangle_{l_\nu}^2\,\rrangle_k,
\end{equation}
and we will call such nucleons \emph{inclusively wounded} (\winc), as opposed to \emph{absorptively wounded} (\wabs).

In analogy with \eqref{eq:averl2} we define $W^{(\winc)}_k$ by the
relation
\begin{equation}
W^{(\winc)}_k (\nucrelpos)\equiv 1-
\llangle \,S^{(\pp)}_{k,l}(\nucrelpos) \,\rrangle_{l_\nu}^2
 =1-\left(1-T_k^{(\pp)}(\nucrelpos)\right)^2,
\label{eq:defY}
\end{equation}
which gives the probability that the target nucleon $\nu$ is either
absorbed or diffractively excited, by a projectile in state $k$. Thus,
if these target nucleons are counted as wounded, the cross section is
also given by \eqref{eq:sigmaabspA}, when $W_k^{(\wabs)}$ is replaced
by $W^{(\winc)}_k$.  We note that the expression for the wounded
nucleon cross section resembles the total one in
\eqsref{eq:targaverage2} and \eq{eq:elgribov}, with $T_k^{(\pp)}$
replaced by $W_k^{(\wabs)}$ or $W_k^{(\winel)}$. Note also that as
$W^{(\winc)}_k$ is determined via \eqref{eq:defY}, when $T_k^{(\pp)}$
is known including its $\mbf{b}$-dependence. This is not the case for
$W^{(\wabs)}_k$, which contains the average over target states of the
\emph{square} of the amplitude $T^{(\pp)}_{k,l}$.

\vspace{2mm}

\textbf{Elastically scattered projectile protons}
\vspace{2mm}

We should note that the probabilities given above include events,
where the projectile is elastically scattered, and thus not regarded
as a wounded nucleon. The probability for this to happen in an event
with diffractively excited target nucleons, is given by the relation
($\langle\cdots\rangle_p$ and $\langle\cdots\rangle_t$ denote averages
over projectile and target states respectively)
\begin{equation}
  \llangle\,\llangle\,S\,\rrangle_{p}^2\,\rrangle_{t} - \left(\llangle
    S\,\rrangle_{p,t}\right)^2 =
  \llangle\, \llangle\, \prod_\nu S_{k,l_\nu}\,\rrangle_k^2\,\rrangle_{l_\nu} -
  \left\{\,\llangle\, \llangle\,\, 
    \prod_\nu S_{k,l_\nu}\,\rrangle_k\,\rrangle_{l_\nu}\,\right\}^2.
\end{equation}
In case these events do not contribute to the observable under study,
this contribution should thus be removed. For a large target nucleus,
this is generally a small contribution.

\subsubsection{Wounded nucleon multiplicity}
\label{sec:woundedmult}

In the following we let $W_k$ denote either $W^{(\winc)}_k$ or
$W^{(\wabs)}_k$, depending upon whether or not diffractively excited
target nucleons should be counted as wounded.

\vspace{2mm}

\textbf{Average number of wounded nucleons}
\vspace{2mm}

As $W_k(\nucrelpos)$ denotes the probability that target nucleon $\nu$
is wounded, the average number of wounded nucleons in the target is
then (for fixed $\mbf{b}$) given by $\langle \sum_\nu
W_k(\nucrelpos)\rangle_{k,\nucpos}$, obtained by summing over target
nucleons $\nu$, and averaging also over projectile states $k$ and all
target positions $\nucpos$. Averaging over impact parameters,
$\mbf{b}$, is only meaningful, when we calculate the average number of
wounded target nucleons per event with at least one wounded nucleon,
which we denote $\langle N\subw^t \rangle$. This is obtained by
dividing by the probability in \eqref{eq:sigmaabspA}. Integrating over
$\mbf{b}$, weighting by the same absorptive probability, and
normalising by the total absorptive cross section (also integrated
over $\mbf{b}$) we get
\begin{equation}
\llangle N\subw^t \rrangle = \frac{\int d^2 b\sum_\nu \llangle\,\llangle\,
  W_{k}(\nucrelpos)  \rrangle_k\,\rrangle_{\nucpos} }
{\int d^2 b\, \llangle\,\llangle\, 1 - \prod_\nu\left(1-
  W_{k}(\nucrelpos)\right) \,
  \rrangle_k\,\rrangle_{\nucpos}}.
\label{eq:dPdNpart}  
\end{equation}
Note that the total number of wounded nucleons is given by
$N\subw=N\subw^t + 1$, as the projectile proton should be added,
provided the projectile proton is not elastically scattered (in which
case all wounded target nucleons have to be diffractively excited).

\vspace{2mm}

\textbf{Multiplicity distribution for wounded nucleons}
\label{sec:woundeddistr}
\vspace{2mm}
 
It is also possible to calculate the probability distribution in the
number of wounded target nucleons $N\subw^t$. For fixed projectile
states $k$ and target nucleon positions $\nucrelpos$, the probability
for target nucleon $\nu$ to be wounded, or not wounded, is
$W_k(\nucrelpos)$ and $1-W_k(\nucrelpos)$ respectively. For
\emph{fixed} $k$ the probability distribution in the number of
absorbed target nucleons is then given by
\begin{equation}
  \frac{d\,P_k(\mbf{b})}{d\,N\subw^t} = \sum_{\mathcal{C}_{N\subw^t}} 
  \prod_{\nu \in \mathcal{C}_{N\subw^t}}  W_k(\nucrelpos)
   \prod_{\mu \in
    \overline{\mathcal{C}}_{N\subw^t}}
  \left\{1-W_k(\mbf{\tilde{b}}_\mu)\right\}.  
\label{eq:woundeddistrib}
\end{equation}
Here the sum goes over all subsets $\mathcal{C}_{N\subw^t}$ of
$\subw^t$ wounded target nucleons, and
$\overline{\mathcal{C}}_{N\subw^t}$ is the set of the remaining
$A-N\subw^t$ target nucleons, which thus are not wounded.  The states
of the target nucleons can be assumed to be uncorrelated, and the
averages could therefore be taken separately, as in
\eqref{eq:averl2}. The state $k$ and positions $\nucpos$ or
$\mbf{b}_\mu$ give, however, correlations between the different
factors, and these averages must be taken after the multiplication,
which gives the result
\begin{equation}
\frac{d\,P(\mbf{b})}{d\,N\subw^t} =
\llangle\,\llangle\,\left\{
\sum_{\mathcal{C}_{N\subw^t}} 
  \prod_{\nu \in \mathcal{C}_{N\subw^t}} 
  W_k(\nucrelpos) \prod_{\mu \in
    \overline{\mathcal{C}}_{N\subw^t}} 
  \left\{1-W_k(\mbf{\tilde{b}}_\mu)\right\}
\right\}
\,\rrangle_{\nucpos}\,\rrangle_k. 
\label{eq:woundeddistaver}
\end{equation} 

The distribution in \eqref{eq:woundeddistaver} includes the
possibility for $N\subw^t=0$. As for the average number of wounded
nucleons above, to get the normalised multiplicity distribution for
events, with $N\subw^t\geq 1$, we should divide by the probability in
\eqref{eq:sigmaabspA}. The final distribution is then obtained by
integrating over $\mbf{b}$, with a weight given by the same absorption
probability. This gives the result
\begin{equation}
\left. \frac{d\,P}{d\,N\subw^t}\right|\sub{ev}=
\frac{\int d^2b\,\,d\,P(\mbf{b})/d\,N\subw^t}
{\int d^2b\,\,d\sigma^{\pA}_{w}(\mbf{b})/d^2b},
\label{eq:Npartdistr}
\end{equation}
where $dP/d\,N\subw^t(\mbf{b})$ and $d\sigma^{\pA}_{w}(\mbf{b})/d^2b$
are the expressions in \eqsref{eq:woundeddistaver}) and
\eq{eq:sigmaabspA}.

We want here to emphasise that the quantity $W^{(\wabs)}_k$ contains
the average of the \emph{square} of the amplitude $T^{(\pp)}$, and is
therefore not simply determined from the average $\langle
T^{(\pp)}\rangle_l= \langle 1-S^{(\pp)}\rangle_l$, which appears in
the expression for the total and elastic cross sections in
\eqsref{eq:targaverage2} and \eq{eq:elgribov}. This contrasts to the
situation for inclusively wounded nucleons, where $W_k^{(\winc)}$ in
\eqref{eq:defY} actually is directly determined by $\langle
T_{k,l}^{(\pp)} \rangle_l$.

\subsection{Nucleus geometry and quasi-elastic scattering}
\label{sec:nucleusgeometry}

In a real nucleus the nucleons are subject to forces with a hard
repulsive core, and their different points $\mbf{r}_\nu$ are therefore
not uncorrelated. In Glauber's original papers this correlation was
neglected, and this approximation is discussed in the subsequent
section.

In addition to the suppression of nucleons at small separations, the
geometrical structure will fluctuate from event to event. These
fluctuations are not only a computational problem, but have also
physical consequences. Just as fluctuations in the nucleon
substructure can induce diffractive excitation of the nucleon,
fluctuations in the nucleus substructure induces diffractive
excitation of the nucleus. If the projectile is elastically scattered
these events are called quasi-elastic. The fluctuations in the target
nucleon positions are also directly reproduced by the Monte Carlo
programs mentioned above, and within the Good--Walker formalism the
quasi-elastic cross section, $\sigma\subqel$, is given by (\cf\
\eqref{eq:sigmadiff}):
\begin{equation}
  d\sigma\subqel/d^2b=
  \llangle\llangle T\rrangle_{p}^2\rrangle_{t}\,\,.
  \label{eq:qel}
\end{equation}
The average over the target states here includes averaging over all
geometric distributions of nucleons in the nucleus, and all partonic
states of these nucleons. Note that this expression includes the
elastic proton--nucleus scattering (given by $\llangle
T\rrangle_{p,t}^2$).  Some results for quasi-elastic \pPb\ collisions
are presented in ref.~\cite{Alvioli:2009ab,Gustafson:2015ada}.

\subsection{Optical limit -- uncorrelated nucleons and large nucleus
  approximations} 
\label{sec:uncorrelated}

Even though the averages in \eqsref{eq:totgribov} and
\eq{eq:sigmaabspA} factorise, they are still complicated by the fact
that all factors $S^{(\pN_\nu)}$ are different, due to the different
values for the impact parameters. It is interesting to study
simplifying approximations, assuming uncorrelated nucleon positions
and large nuclei. This is generally called the optical limit. It was
used by Glauber in his initial study \cite{Glauber:1955qq}, and is
also described in the review by Miller \emph{et al.}
\cite{Miller:2007ri}, for a situation when diffractive excitation is
neglected. We here discuss the modifications necessary when
diffractive excitation is included, also separating single excitation
of projectile and target, and double diffraction.

\subsubsection{Uncorrelated nucleons}

Neglecting the correlations between the nucleon positions in the
target nucleus, the individual nucleons can be described by a smooth
density $A\cdot\rho(b_\nu)$ (normalised so that $\int d^2b\, \rho(b) =
1$).  In this approximation all factors $\llangle
S^{(\pN_\nu)}_{k,l_\nu}\rrangle_{t} = 1 - \llangle
T^{(\pN_\nu)}_{k,l_\nu}\rrangle_{t} $ in \eqref{eq:targaverage}, which
enter the total \pA\ cross section in \eqref{eq:totgribov}, are
uncorrelated and give the same result, depending only on projectile
state and impact parameter $k$ and $\mbf{b}$:
\begin{equation}
\llangle T^{(\pN_\nu)}_{k,l_\nu}(\mbf{b} - \nucpos)\rrangle_{t}=
\int d^2 b_\nu \,\rho(\nucpos) \llangle
T^{(\pp)}_{k,l}(\mbf{b}-\nucpos)\rrangle_{l}.
\label{eq:uncorrelated1}
\end{equation}

In the same way all factors $W_k(\nucrelpos)$,  entering the
wounded nucleon cross sections in \eqsref{eq:averl2} and \eq{eq:defY}, give
equal contributions: 
\begin{eqnarray}
  \llangle W_k^{(\wabs)}(\nucrelpos )\rrangle_{\nucpos}=
  \int d^2 b_\nu \,\rho(\nucpos) \left(1-\llangle \left( S_{k,l}^{(\pp)}(\mbf{b} -
      \nucpos)\right)^2 \rrangle_{l} \right);\nonumber\\
  \llangle W_k^{(\winc)}(\nucrelpos )\rrangle_{\nucpos}=
  \int d^2 b_\nu \,\rho(\nucpos)  \left(1-\llangle S_{k,l}^{(\pp)}(\mbf{b} -
    \nucpos) \rrangle_{l}^2 \right).
  \label{eq:uncorrelated3}
\end{eqnarray}

\subsubsection{Large nucleus}

If, in addition to the approximations in \eqsref{eq:uncorrelated1} and
\eq{eq:uncorrelated3}, the width of the nucleus (specified by $\rho$)
is much larger than the extension of the \pp\ interaction (specified
by $T^{(\pp)}$), further simplifications are possible. For the
amplitude in \eqref{eq:uncorrelated1} we can integrate over $\nucpos$,
and get the approximation
\begin{equation}
\llangle T^{(\pN_\nu)}_{k,l_\nu}(\mbf{b} - \nucpos)\rrangle_{t} \approx
\rho(\mbf{b}) \int d^2 \tilde{b}\, \llangle
T^{(\pp)}_{k,l}(\tilde{b})\rrangle_{l} =
\rho(\mbf{b})\,\sigma^{\pp}\subtoti{k}\, /2.
\label{eq:uncorrelated2} 
\end{equation}
We have here introduced the notation $\sigma^{\pp}\subtoti{k}$ for the
total cross section for a projectile proton in state $k$, averaged
over all states for a target proton.

In the same way we get
\begin{equation}
\llangle W_{k}(\nucrelpos)\rrangle_{t} \approx
\rho(\mbf{b}) \int d^2 \tilde{b}\, W_k(\tilde{b}) =
\rho(\mbf{b})\,\sigma^{\pp}_{w,k},
\label{eq:Woptical}
\end{equation}
where $W_k$ is either $W_k^{(\wabs)}$ or $W_k^{(\winc)}$, and
$\sigma^{\pp}_{w,k}$ is the corresponding \pp\ cross section for a
projectile in state $k$.

\subsubsection{Total cross section}

Inserting \eqref{eq:uncorrelated2} into \eqsref{eq:totgribov} -
\eq{eq:elgribov} gives the total cross section for a projectile in
state $k$ hitting a nucleus:
\begin{eqnarray}
d\sigma^{(\pA)}\subtoti{k}/d^2b&=& 2\llangle T^{(\pA)}_{k,l}(\mbf{b})\rrangle_{t} =
2\left\{1- \left(1- \rho(\mbf{b}) \, \sigma^{\pp}\subtoti{k}\,/2  \right)^A
\right\}\nonumber=\\
&=& -2\sum_{N=1}^A \binom{A}{N}
\left(-\, \rho(b)\,\sigma^{\pp}\subtoti{k}/2\, \right)^N.
\label{eq:TpAlinear}
\end{eqnarray}
The total \pA\ cross section is then finally obtained by averaging
over projectile states, $k$, and integrating over impact parameters,
$\mbf{b}$:
\begin{equation}
\sigma^{(\pA)}\subtot= \int d^2 b\, \llangle\, d\sigma^{(\pA)}\subtoti{k}/d^2b
\,\rrangle_k. 
\label{eq:sigmatotlinear} 
\end{equation}

We note here in particular, that in this approximation the
$b$-dependence of $T^{(\pp)}_k(b)$ is unimportant, and the result
depends only on its integral $\sigma^{\pp}\subtoti{k}/2$. We also note
that to calculate the elastic \pA\ cross section $\sim \int d^2b\,
(T(b))^2$, which has a steeper $b$-dependence, a knowledge about this
dependence is also needed.

\emph{Proton-deuteron cross section}

Neglecting fluctuations, \eqsref{eq:TpAlinear} and
\eq{eq:sigmatotlinear} would give the simpler result
\begin{equation}
\sigma^{(\pA)}\subtot=  -2\sum_{N=1}^A \binom{A}{N}
\left(\left(-\frac{\sigma\subtot}{2}\right)^N \int d^2b\, \rho^N(b)
\right). 
\label{eq:sigmatotoptical} 
\end{equation}
For the special case with a deuteron target we then get the
result\footnote{Although the deuteron has only 2 nucleons, it is very
  weakly bound, and its wave function is extended out to more than
  5~fm. Therefore the large nucleus approximation is meaningful also
  here.}
\begin{equation}
\sigma^{\pd}\subtot=2 \sigma\subtot^{\pp} - \frac{1}{2}\left( \int d^2b
\,\rho^2(b)\right)\, 
\left(\sigma\subtot^{\pp}\right)^2,
\label{eq:pdnofluc}
\end{equation}
and with the estimate $\int d^2b \,\rho^2(b) = 1/(2\pi \langle b^2
\rangle)$ describing the deuteron wavefunction, we recognise Glauber's
original result.

For a non-fluctuating amplitude, the optical theorem gives a direct
connection between the total and elastic cross sections. As the
integral over $d^2b$ gives the Fourier transform at $q=0$, we have
\begin{equation}
\sigma\subtot^{\pp}=
2\int d^2b\,  T^{(\pp)}_{k,l}(b) =4\pi \tilde{T}^{(\pp)}_{k,l}(q=0)=\sqrt{16\pi
\left.\frac{d}{dt} \sigma^{\pp}\subel(t)\right|_{t=0}}.
\end{equation}
Here $\tilde{T}(q)$ denotes the Fourier transform of the amplitude $T(b)$. 
For a Gaussian interaction profile we get
\begin{equation}
(\sigma\subtot^{(\pp)})^2 \propto
\sigma^{\pp}\subel\cdot B,
\label{eq:sigmaelB}
\end{equation} 
where the slope $B$ is a measure of the width of the interaction.  As
$\sigma\subel$ is determined by the squared amplitude, the ratio
$\sigma\subel/\sigma\subtot$ will be larger for a strong interaction
with a short range, than for a weaker interaction with a wider range.

For the general case with fluctuating amplitudes, we can using the
results in \eqref{eq:sigmadiff}, in an analogous way rewrite
$(\sigma^{(\pp)}\subtoti{k})^2$ in \eqref{eq:TpAlinear} in the
following form
\begin{equation}
\left(\sigma^{(\pp)}\subtoti{k}\right)^2= 16\pi^2 \langle \langle
\tilde{T}^{(\pp)}_{k,l}(q=0) \rangle_l^2 \rangle_k = 16\pi
 \left.\frac{d}{dt} \left( \sigma^{\pp}\subel(t) +
\sigma^{\pp}\subsdp(t)\right)\right|_{t=0}.
\label{eq:averTlsquare} 
\end{equation}
Here $\sigma^{\pp}\subsdp$ denotes the cross section for single
diffractive excitation of the projectile proton (\ie\ on one side
only).  For a fluctuating amplitude we then get instead of
\eqref{eq:pdnofluc}
\begin{equation}
\sigma^{\pd}\subtot=
2\, \sigma^{_(\pp)}\subtot - 8\pi \left( \int d^2 b\,
  \rho^2(b) \right)  
\left.\frac{d}{dt} \left( \sigma\subel^{\pp}(t) +
\sigma^{\pp}\subsdp(t)\right)\right|_{t=0}.
\label{eq:deuteriumfluct}
\end{equation}

The negative term in \eqref{eq:deuteriumfluct} represents a shadowing
effect, which for a deuteron target has one contribution from the
elastic proton--nucleon cross section, and another from diffractive
excitation. Note in particular, that it is only \emph{single}
diffraction which enters, with an excited projectile but an
elastically scattered target nucleon. (This would be particularly
important in case of a photon or a pion projectile.)

\emph{Larger target nuclei} For a larger target higher moments,
$\llangle \llangle T^{(\pp)}\rrangle_t^n\rrangle_p$ ($n=$ 1, 2,
\ldots, A), of the \pp\ amplitude, averaged over target states, are
needed. These moments cannot be determined from the total cross
section and the cross section for diffractive excitation.  They can be
calculated if we know the full probability distribution, $dP/d\llangle
T^{(\pp)} \rrangle_t$, for the \pp\ amplitude averaged over target
states, but for varying projectile states\footnote{The average for
  $n=3$ was estimated from diffractive proton-deuteron scattering in
  ref.~\cite{Blaettel:1993ah}.}.  In addition also higher moments of
the nucleus density, $\int d^2 b\, \rho^n(b)$, are needed.

We also note here that the factorisation feature in \eqref{eq:glauber}
is not realised in \AA\ collisions. This implies that also in the
optical limit, the \AA-results cannot be directly expressed in terms
of the moments $\llangle \llangle T^{(\pp)}\rrangle_t^n\rrangle_p$.

\subsubsection{Wounded nucleon cross sections}

Also for cross sections corresponding to wounded (absorptively or
inclusively) nucleons,
approximations analogous to \eqsref{eq:uncorrelated1} and
\eq{eq:uncorrelated2} are possible.
Integrating the expressions in \eqref{eq:Woptical} over
$\nucpos$, and averaging 
also over projectile states $k$ gives, in analogy with
\eqsref{eq:TpAlinear} and \eq{eq:sigmatotlinear}, the following result
\begin{equation}
d\sigma^{\pA}\subw/d^2b=1 
-\llangle \,\left(1-\rho(b)\,\sigma^{\pp}_{w,k}\right)^A \,\rrangle_k.
\label{eq:absuncorr}
\end{equation}
The average in \eqref{eq:absuncorr} includes averages of all possible
powers $\llangle(\sigma^{\pp}_{w,k})^n\rrangle_k$. For $n=1$ this is
just equal to the \pp\ cross section $\sigma^{\pp}\subw$ for (with $w$
denoting either absorptively or inclusively wounded), but for higher
moments a knowledge of the full probability distribution for
$\sigma^{\pp}_{w,k}$ is needed, in analogy with
\eqref{eq:sigmatotlinear} for the total \pA\ cross section.  Note,
however, that a similar relation is not satisfied for the elastic or
total inelastic cross sections, $\sigma\subel$ and $\sigma\subineltot
= \sigma\subtot - \sigma\subel$, which as seen in \eqref{eq:sigmadiff}
contain the average over projectile states $k$ before squaring.

\subsubsection{Average number of wounded nucleons}

In \eqref{eq:absuncorr} $\rho(b)\sigma^{\pp}_{w,k}$ represents the
probability that a specific target nucleon is wounded, in a collision
with a projectile in state $k$ at an impact parameter $\mbf{b}$. In
the optical limit this probability is the same for all $A$ target
nucleons. Averaging over projectile states $k$ then gives the average
number of wounded target nucleons for an encounter at this
$b$-value. Dividing by the probability for a ``wounded'' event, we get
the average number of wounded target nucleons per wounded event for
this $b$:
\begin{equation}
\llangle\,N\subw^{t}(b)\,\rrangle = \frac{A\,\rho(b) 
\,\sigma^{\pp}\subw}
{1 -\llangle \,\left(1-\rho(b)\sigma^{\pp}_{w,k}\right)^A \,\rrangle_k}.
\label{eq:Npartuncorr} 
\end{equation}
Normalising by the probability for absorption in \eqref{eq:absuncorr},
and integrating over $b$ with a weight given by the same probability,
then gives
\begin{equation}
\llangle\,N\subw^{t}\,\rrangle=
 \frac{\int d^2 b\, A\,\rho(b) \,\sigma^{\pp}\subw}
  {\int d^2 b\, d\sigma^{\pA}\subw/d^2b},
\end{equation}
with $ d\sigma^{\pA}\subw/d^2b$ given by \eqref{eq:absuncorr}.  As
noted above, this needs knowledge of the full probability distribution
for $\sigma^{\pp}_{w,k}$.

\subsubsection{Multiplicity distribution for wounded nucleons}

As in \sectref{sec:wounded},
when calculating the full distribution in $N\subw^{t}(b)$,
it is important to take the average over projectile states $k$ after
multiplication of the different nucleon absorption probabilities, which gives
\begin{equation}
\frac{dP(b)}{d\,N\subw^{t}} = \binom{A}{N\subw^{t}} \llangle\,\left(\rho(b)\,\sigma^{\pp}_{w,k}\right)^{N\subw^{t}}
\cdot \left(1- \rho(b)\,\sigma^{\pp}_{w,k}\right)^{A-N\subw^{t}}\,\rrangle_k.
\label{eq:Ndistb}
\end{equation} 
Similar to the general result in \eqref{eq:woundeddistaver}, this
expression includes the probability for zero target participants.
Normalising by the probability for absorption in \eqref{eq:absuncorr},
and integrating over $b$ with a weight given by the same probability,
gives finally
\begin{equation}
\frac{dP}{d\,N\subw^{t}} = \frac{\int d^2 b\,\,
  dP(b)/d\,N\subw^{t}}{\int d^2 b\,\, d\sigma^{\pA}\subw/d^2b}.
\label{eq:Ndist}
\end{equation}
Here $dP(b)/d\,N\subw^{t}$ and $d\sigma^{\pA}\subw/d^2b$ are the
expressions in \eqsref{eq:Ndistb} and \eq{eq:absuncorr}.

\section{Models for \pp\ scattering used in Glauber calculations}
\label{sec:models}

As mentioned in \sectref{sec:nucleusgeometry}, most analyses today use
a Monte Carlo simulation to generate a realistic distribution of
nucleons within the nucleus, including fluctuations which cause
quasi-elastic scattering of the nucleus
\cite{Alvioli:2009ab,Gustafson:2015ada} as well as initial state
anisotropies (\eg\ \cite{Alvioli:2011sk}). In contrast most Glauber
Monte Carlos use a rather simple model for the \pp\ interaction. In
this section we discuss some models which have been used in analyses
of experimental data. We will also comment on the pros and cons, when
these models are applied to \pA\ collisions.

In the optical approximation, where the extension of the nucleus is
much larger than the range of the \pp\ interaction, the results for
\pA\ collisions can be expressed in terms of integrated \pp\
amplitudes, without knowledge of their respective impact parameter
dependence (see \eqref{eq:uncorrelated2}). It is therefore most
essential to use a model, where the integrated \pp\ cross sections are
well reproduced.  Note, however, that although the total \pA\ cross
section is most sensitive to the integrated total \pp\ cross section,
the $b$-dependence is very important for the ratio between the elastic
and total cross sections (see \eqref{eq:sigmaelB}). This feature
naturally also affects the ratio between the inelastic and the total
cross sections.

As mentioned in the introduction, the problems encountered when taking
fluctuations and diffractive excitation of the nucleons properly into
account in the Glauber model, have implied that these effects are
neglected or severely approximated in many applications, see \eg\
ref.~\cite{Miller:2007ri}. However also in models which do include
fluctuations, as far as we know no published analysis uses a model
which can separate single excitation of the projectile from that of
the target, and from double excitation.  This is a problem as the
various \pA\ cross sections in \sectref{sec:glauber} contain powers of
\pp\ amplitudes averaged in different ways over projectile and target
fluctuation.

We first discuss some simple models determined by just a few
parameters, and then the more ambitious approach by Strikman and
coworkers, using a continuous distribution for the fluctuations.

\subsection{Simple approximations}
\label{sec:simple-models}

\subsubsection{Non-fluctuating models}

\textbf{i) Black disk model}
\vspace{2mm}

The simplest approximation is the ``black disk model'' with a fixed
radius. Here diffractive excitation is completely neglected, and the
target in a nucleon--nucleon collision acts as a black absorber. The
projectile nucleon travels along a straight line, and interacts
inelastically if the transverse distance to a nucleon in the target is
smaller than a distance $R$, which gives
\begin{equation}
T^{(\pp)}(b)=\Theta(R-b)
\label{eq:blackdisk}
\end{equation}
This results in the following cross sections:
\begin{equation}
\sigma\subel = \sigma\subineltot =\sigma\subtot/2=\pi
R^2,\,\,\,\,\,\,\sigma_{D}=0.
\end{equation} 
Here $\sigma\subdtot$ denotes the cross section for diffractive
excitation.  (See \eqref{eq:sigmael-tot} with $F=\infty$.) This is in
clear contrast to the experimental result $\sigma\subel\approx
\sigma\subtot/4$ and the total diffractive excitation of the same
order of magnitude as $\sigma\subel$.  This again illustrates how a
short range amplitude gives a large $\sigma\subel/\sigma\subtot$
ratio.  The radius can therefore be adjusted to reproduce the
experimental value for one of these three cross sections, at the cost
of not reproducing the other two.

As discussed in ref.~\cite{Gustafson:2015ada}, choosing to reproduce
$\sigma\subtot^{\pp}$, the simple black-disk result for \pPb\
collisions agrees rather well with the \dipsy model for
$\sigma\subtot^{\pPb}$, but not so well for $\sigma\subel^{\pPb}$ or
$\sigma\subineltot^{\pPb}$. Similarly adjusting $R$ to reproduce
$\sigma\subineltot^{\pp}$ or $\sigma\subabs^{\pp}$ gives results which
agree with \dipsy for the corresponding \pPb\ cross section, but not
for the other.

The black disk model is implemented in many Monte Carlos, \eg\ in the
\phobos Monte Carlo \cite{Alver:2008aq,Loizides:2014vua}. It is also
used in refs.~\cite{Alvioli:2009ab, Alvioli:2011sk} where the authors
study fluctuations in the distribution of nucleons within the nucleus,
but do not address the fluctuations in the \pp\ interaction.

\vspace{2mm}

\textbf{ii) Grey disk and Gaussian profile}
\vspace{2mm}

Also other shapes for a non-fluctuating \pp\ interaction have been
used in the literature \cite{Ding:1989mj,Pi:1992ug,Miller:2007ri}. The
simplest example is a fixed semi-transparent ``grey disk'', with
opacity given by the parameter $\alpha$:
\begin{equation}
T^{(\pp)}(b)=\alpha \Theta(R-b),
\end{equation}
which gives $\sigma\subel :\sigma\subtot: \sigma\subineltot =
\alpha:2:2-\alpha$.

Another example is a Gaussian profile 
\begin{equation}
T^{(\pp)}(b)=\alpha\, \exp(-b^2/2B)
\label{eq:gaussianprofile}
\end{equation}
giving $\sigma\subel :\sigma\subtot: \sigma\subineltot =
\alpha:4:4-\alpha$.

These models contain two parameters (with $\alpha\leq 1$ to satisfy
the unitarity constraint $T\leq1$), and it is therefore possible to
fit \eg\ the total and the elastic cross sections, with the inelastic
(non-diffractive) cross section given by the difference between these
two. The lower ratio $\sigma\subel/\sigma\subtot$ is a consequence of
the wider interaction range.  We note, however, that even if typical
events are well reproduced it is often interesting to study rare
events in the tail of a distribution. As an example the tail of the
\pp\ amplitude out to large $b$-values may be important for the
probability to produce rare events with many \pN\ sub-collisions at
large impact separation. The Gaussian profile may \eg\ thus give a
larger tail than the gray disk, also when they give very similar
averages.

\subsubsection{Models including fluctuations}

To account for diffractive excitation, we must allow the \pp\
amplitude to fluctuate. Models used in the literature do, however, not
separate fluctuations in the projectile and the target. From
\eqsref{eq:targaverage2} and \eq{eq:elgribov} we see that if the
amplitude is adjusted to reproduce the amplitude averaged over target
states, $\langle T^{(\pp)}(b) \rangle_t$, then the correct result for
the \pA\ total cross section will be obtained. The fluctuations
included in the model should then only describe fluctuations in the
projectile state, and should thus reproduce the cross section for
single excitation of the projectile. Such a model will, however, not
reproduce cross sections for absorptively wounded nucleons properly,
as will be discussed further below.  \vspace{2mm} \pagebreak[3]
\textbf{iii) Fluctuating grey disk} \vspace{2mm}

The simplest model accounting for diffractive excitation is the fluctuating
``grey disk model''.
Here it is assumed that within a radius $R$ the projectile is absorbed
with probability $a$, with $0<a<1$.  This implies that $\langle
T(b)^2\rangle=\langle T(b)\rangle$, and the resulting \pp\ cross
sections are here
\begin{eqnarray}
  \sigma\subtot    &=& 2 \int d^2 b \llangle T^{(\pp)}(b) \rrangle 
  = 2 \pi R^2 a                  \nonumber \\
  \sigma\subel     &=&   \int d^2 b \llangle T^{(\pp)}(b) \rrangle^2
  =   \pi R^2 a^2                 \nonumber \\
  \sigma\subdtot
  &=&   \int d^2 b \left(\llangle T^{(\pp)}(b)^2 \rrangle -
    \llangle T^{(\pp)}(b) \rrangle^2\right) = \pi R^2 a(1-a) \nonumber \\
  \sigma\subabs &=&  
  \int d^2 b \llangle 1 - \left(1 - T^{(\pp)}(b)\right)^2\rrangle
  =   \pi R^2 a.
\label{eq:greydisk}
\end{eqnarray}
The two parameters $R$ and $a$ can now be adjusted to reproduce \eg\
the total and the elastic \pp\ cross sections. At LHC this would give
$a\approx 1/2$. The cross section for diffractive excitation should
here be interpreted as representing only the single excitation of the
projectile, while target excitation is part of the absorptive cross
section. With $a=1/2$ this is quite an overestimate. It corresponds
rather to the total diffractive excitation, which implies that the
results for $\sigma\subabs/\sigma\subtot$ is close to the experimental
value.  The relation between the absorptive and diffractive cross
section, which together make up the inelastic cross section, is also
fixed in this model.

The agreement of the fluctuating gray disk with \dipsy results for \pPb\ collisions are not superior to 
those of the black disk model \cite{Gustafson:2015ada}.
\vspace{2mm}

\textbf{iv) Fluctuating Gaussian profile}
\vspace{2mm}

Here the profile in \eqref{eq:gaussianprofile} gives the probability
for absorption. Thus $T=1$ with probability $\alpha\, \exp(-b^2/2B)$
while $T=0$ with probability $1-\alpha\, \exp(-b^2/2B)$. As for the
fluctuating gray disk this implies that $\langle T(b)^2\rangle=\langle
T(b)\rangle$. This does not change the total and elastic cross
sections, but it splits the inelastic one into relative fractions a
non-diffractive (absorptive) and a diffractive part, with the result
$\sigma\subtot: \sigma\subel: \sigma\subdtot :\sigma\subabs =
4:\alpha:2-\alpha:2$. For $\alpha\approx 1$ this gives the same result
as the fluctuating gray disk. (Although this implies that the results
will be very similar in the optical limit, it does not mean that the
results are identical for a more realistic nucleus. This will be
particularly true for the tail at very large numbers of wounded
nucleons.)

This model is also an option in the \phenix Monte Carlo.
\vspace{2mm}

\textbf{v) Fluctuating black disk model}

\vspace{2mm}

It is possible to let the radius of the black disk fluctuate. As for
the two previous model, the fact that $T(b)$ is either 1 or 0 implies
that $\langle T(b)^2\rangle=\langle T(b)\rangle$, which gives
$\sigma\subabs=\sigma\subtot/2$. Such a fluctuating black disk model
has sometimes been used in connection with the GG model described in
\sectref{sec:Strikman}, and will be further described below.

\vspace{2mm}

\textbf{vi) A new simple model allowing for separate projectile and target
  excitations} 
\vspace{2mm}

The main reason neither of the above models are able to properly take
into account the diffractive aspects of nucleon collisions, is that
the fluctuations in the cross sections are not treated in terms of
fluctuations in the projectile and target separately.  Interpreting
the amplitude $T$ as the average over target states, which as
mentioned above can give a correct total cross section, excitation of
target nucleons will not be separated from the absorptive cross
section.

To redeem this we have constructed a new model, which in some sense is
the minimal possible extension needed to reproduce all relevant
semi-inclusive cross sections. The basis of the model is having
fluctuating sizes of the colliding nucleons. With some probability,
$c$, a nucleon with radius $r_1$, can fluctuate into a larger radius
$r_2$. This will then give us the elastic amplitude for a projectile
with radius $R_p$ colliding with a target with radius, $R_t$,
\begin{equation}
  \label{eq:newmod}
  T(b)=\alpha\,\Theta(R_p+R_t-b).
\end{equation}
Here $\alpha$ is again an opacity parameter between zero and one,
which together with $c$, $r_1$ and $r_2$ gives us four parameters
which can be adjusted to reproduce the relevant nucleon--nucleon cross
sections $\sigma\subabs$, $\sigma\subel$,
$\sigma\subsdp=\sigma\subsdt$ and $\sigma\subdd$. Below we will refer
to this model as \newdisk.

\subsection{The approach by Strikman and coworkers}
\label{sec:Strikman}

An ambitious approach to describe fluctuations in \pp\ scattering, for
use in the Glauber model, was presented in
refs.~\cite{Heiselberg:1991is, Blaettel:1993ah}. This model has been
further extended in several papers by Alvioli, Strikman and coworkers;
for a general overview see ref.~\cite{Alvioli:2013vk} and further
references in there. Recent studies, with applications to the LHC,
discuss effects of colour fluctuations (or flickering)
\cite{Alvioli:2014sba}, and evidence for $x$-dependent proton colour
fluctuations \cite{Alvioli:2014eda}.  The model does not take into
account the possibility of separate excitations of the projectile and
the target, and the fluctuations in the target are not considered. In
this section we will also discuss how the model can be modified to
take the target fluctuations into account. (Note that our amplitude
$T$ is in ref.~\cite{Alvioli:2013vk} denoted $\Gamma$.)  \vspace{2mm}

\subsubsection{\boldmath\pp\ total cross section}

The basic feature of the model is a description of the fluctuations in
the \NN\ total cross section, as a smooth function, which has the form
\begin{eqnarray}
P\subtot(\sigma)&=&\rho\, \frac{\sigma}{\sigma + \sigma_0} 
\exp\left\{-\frac{(\sigma/\sigma_0 - 1)^2}{\Omega^2}\right\}, 
\label{eq:Psigmatot}\\ 
\sigma\subtot&=&\int d\sigma\,\sigma\, P\subtot(\sigma).
\label{eq:Psigmatot2}
\end{eqnarray}
Here $\sigma$ is regarded as the total \pp\ cross section in a single
event, with the probability distribution $P\subtot(\sigma)$, while the
observed total cross section $\sigma\subtot$ is given by the average
in \eqref{eq:Psigmatot2}.  \footnote{Note, however, that in
  ref.~\cite{Alvioli:2013vk} the notation is changed, such that
  $\sigma \rightarrow \sigma\subtot$ and $\sigma\subtot \rightarrow
  \sigma\subtot^{(\pp)}$.} For the functional form in
\eqref{eq:Psigmatot}, the average and the width of the distribution
are related to (but not identical to) the parameters $\sigma_0$ and
$\Omega$, while $\rho$ is a normalisation constant.

In \eqsref{eq:elgribov} and \eq{eq:targaverage2} we see that the total
\pA\ cross section contains all possible moments with respect to the
fluctuations in the \emph{projectile} state, but only the average (the
first moment) with respect to the fluctuations in the \emph{target}
state.  Thus, although target fluctuations are not considered
explicitely, we conclude that if $\sigma$ is interpreted as the
\emph{average over target states}
\begin{equation}
\sigma=2 \int d^2b\langle T^{(\pp)}_{k,l}\rangle_l,
\label{eq:sigmastrikman}
\end{equation}
and only the average over \emph{projectile states} is described by the
distribution in \eqref{eq:Psigmatot}, then the total \pA\ cross sections will
(in the large nucleus approximation) be determined in terms of all possible
moments $\langle \sigma^N \rangle$, obtained from the distribution
$P\subtot(\sigma)$, and the average in \eqref{eq:Psigmatot2}
will correctly give the total \pp\ cross section.

With this interpretation the width of the distribution can also be
determined from \eqref{eq:averTlsquare}, which gives the second moment
\begin{equation}
\langle \sigma^2 \rangle = 16 \pi^2 
\langle \langle \tilde{T}^{(\pp)}_{k,l}(t=0)\rangle_l^2\rangle_k =
16\pi \left. \frac{d}{dt} \left(\sigma\subel^{\pp}(t) +
\sigma^{\pp}\subsdp(t)\right)\right|_{t=0}.
\label{eq:avsigma2}
\end{equation}
Here the first term in the parenthesis would give $\langle
\sigma\rangle^2$ corresponding to Glauber's result, while the second,
determined by \emph{single} excitation of the projectile, is the
result of fluctuations in the projectile state.

\Eqref{eq:avsigma2} has been used by Blaettel \emph{et
  al.}~\cite{Blaettel:1993ah} together with \eqref{eq:deuteriumfluct}
to estimate the width from shadowing in \pd\ collisions at fixed
target energies. They also estimated the width from diffractive
excitation data at the CERN \pbarp\ collider.  With data from
\totem~\cite{Antchev:2011vs,Antchev:2013gaa,Antchev:2011zz} and
\alice~\cite{Abelev:2012sea} for elastic and single diffractive cross
sections and elastic forward slope, supplemented by the assumption
that the diffractive slope is approximately half the elastic (as is
the case at 560 GeV \cite{Bernard:1986yh}), we get for 7 TeV the
estimated width $\sqrt{\langle \sigma^2\rangle - \langle \sigma
  \rangle^2} \approx 0.4 \langle \sigma\rangle$.  As mentioned above,
the amplitude for larger nuclei the amplitude in \eqref{eq:TpAlinear}
contains also higher moments of the \pp\ amplitude. Blaettel \emph{et
  al.} estimated also the third moment, $\llangle \sigma^3 \rrangle$
from data for diffractive excitation in \pd\ scattering, and they also
studied other analytic forms. Most recent applications use, however,
the form in \eqref{eq:Psigmatot}, in which the higher moments are
fixed by a determination of the width.  \vspace{2mm}

\subsubsection{Elastic cross section}

We use the notation
\begin{equation}
T^{(\pp)}(b,\sigma)\equiv \llangle T^{(\pp)}_{k,l}(b) \rrangle_l
\label{eq:deft}
\end{equation}
 to describe the
$b$-dependence of the fluctuating cross section $\sigma$ in
\eqref{eq:sigmastrikman}. This gives
\begin{eqnarray}
\sigma &=& \int d^2b\,2\, T^{(\pp)}(b,\sigma)\nonumber\\
d\sigma\subtot / d^2b &=& \int d\sigma\,
P\subtot(\sigma)\,2\, T^{(\pp)}(b,\sigma),\nonumber\\ 
d\sigma\subel / d^2b &=& \left|\int d\sigma\, P\subtot(\sigma)\,
T^{(\pp)}(b,\sigma)\right|^2.
\label{eq:Tbsigma}
\end{eqnarray}
As pointed out earlier, the relation between $\sigma\subel$ and
$\sigma\subtot$ depends on the width of the interaction.  Thus,
although the elastic and total cross sections for fixed $b$ are given
by the same average over target fluctuations, the elastic cross
section is not determined by the $\sigma$-distribution in
\eqref{eq:Psigmatot}, unless it is supplemented by a knowledge of the
$b$-dependence (for all values of $\sigma$).

We here note that the distribution in \eqref{eq:Psigmatot} has a tail
out to large cross sections. The unitarity constraint $T(b)<1$, or
$d\sigma\subtot/d^2b <2$, therefore implies that a large value for
$\sigma$ must be associated with a wider $b$-distribution. The effect
of different assumptions about the $b$-dependence will be discussed in
\sectref{sec:GGTdep}.

This feature implies of course that also the inelastic cross section
cannot be directly determined from \eqref{eq:Psigmatot}.

\vspace{2mm}

\subsubsection{Wounded nucleon cross section}
\label{sec:strikdiff}

As discussed in \sectref{sec:woundedcross}, the definition of a
wounded nucleon may depend upon the specific observables under
consideration.  As pointed out earlier, in cases where the
\textit{absorptive} cross section is the most relevant, this is given
by
\begin{equation}
d\sigma\subabs/d^2 b= 2\,\langle T(b)\rangle_{p,t} - \langle T(b)^2\rangle_{p,t},
\end{equation}
which cannot be determined without knowing how the separate fluctuations in
the projectile and target result in single and double diffractive excitation.
We see that in contrast to the expressions entering the total and elastic
cross sections in \eqref{eq:Tbsigma}, this expression contains also the second
moment with respect to the target fluctuations.

In ref.~\cite{Alvioli:2013vk} Alvioli and Strikman identify the
differential wounded nucleon cross section with the total inelastic
\pp\ cross section (which includes diffractive excitation). In the
hypothetical situation where the target did not fluctuate, after
averaging over projectile fluctuations this also gives the absorptive
(inelastic non-diffractive) cross section.

However, if $T$ is identified with the amplitude averaged over target
states, as in \eqref{eq:deft} (which gives the correct result for the
total cross section), then we get instead
\begin{equation}
  \llangle 2T(b)-T(b)^2 \rrangle_p= \llangle T^{(\pp)}_{k,l}(b) \rrangle_{l,k} -
  \llangle \left(\llangle T^{(\pp)}_{k,l}(b) \rrangle_l\right)^2 \rrangle_k
  = d\sigma\subw/d^2b.
\label{eq:strikwounded}
\end{equation}
From \eqref{eq:sigmadiff} we see that this corresponds exactly to the
inclusively wounded nucleon cross section $d\sigma\subwinc/d^2b$,
where $\sigma\subwinc$ now includes diffractively excited target
nucleons:
\begin{equation}
  \label{eq:woundedcross}
  \sigma\subwinc=\sigma\subabs + \sigma\subdd + \sigma\subsdt = \sigma\subtot -
  \sigma\subel - \sigma\subsdp. 
\end{equation}
We here note that, although $d\sigma\subw/d^2b$ in
\eqref{eq:strikwounded} contains the same average of
$T^{(\pp)}_{k,l}(b)$ over target states, to integrate this expression
over $b$ we also need to know the $b$-dependence of
$T^{(\pp)}(b,\sigma)$ for all $\sigma$. We also note that the integral
$\int d^2 b \int d\sigma P(\sigma) T^2(b,\sigma)$ appearing in
$\sigma\subw$ is different from $\int d^2 b [\int d\sigma P(\sigma)
T(b,\sigma)]^2$ appearing in $\sigma\subel$.

The distribution $P\subw(\sigma\subw)$ is consequently not easily
related to the distribution in the total cross section
$P\subtot(\sigma\subtot)$. Lacking a detailed description, Strikman
\textit{et al.} use an approximation assuming the proportional
distribution which for the absorption probability would mean
\begin{equation}
P\subabs(\sigma) \propto P\subtot(\sigma/\lambda\subabs),
\label{eq:Psigmaabs}
\end{equation}
where $\lambda\subabs=\sigma\subabs/\sigma\subtot$.  This
approximation may be less accurate, since for non-peripheral
collisions $T(b)$ is rather close to 1, where $d
P\subabs(b)/dT(b)\equiv d(2T(b) -T(b)^2)/dT(b)=0$, while for
peripheral collisions with small $T$ we have $d P\subabs(b)/dT(b)=2$.
Also, even if the analytic form in \eqref{eq:Psigmatot} may give a
satisfying result, there is no obvious reason why the same value of
the width parameter $\Omega$, should be applicable as the one
determined from shadowing or diffractive excitation.

\textbf{Monte Carlo implementations}

The GG model has been implemented in Monte Carlo simulations in many
applications to \pA\ collisions, \eg\ in ref.~\cite{Alvioli:2013vk,
  Alvioli:2014sba, Alvioli:2014eda}. In experimental analyses it has
been combined with earlier Monte Carlos, where the parameters in one
of the simple models described in \sectref{sec:simple-models} are
allowed to vary according to \eqref{eq:Psigmatot} (or using a scaled
version as in \eqref{eq:Psigmaabs}, but typically using the total
inelastic cross section rather than the absorptive), in a way
reproducing the total (or the inelastic) cross section
respectively. The \phobos\ Monte Carlo \cite{Alver:2008aq} with a
black disk with a variable radius, which is also used by \eg\ \atlas
\cite{Aad:2015zza}.  Fitting to the inelastic \pp\ cross section here
overestimates the number of absorbed nucleons. In
ref.~\cite{Alvioli:2014sba} it is argued that this is a small effect,
as the cross section for diffractive excitation of the projectile
proton is small in \pA\ collisions. However, the cross section for
target nucleon excitation is not small, and although the cross section
for diffractive projectile excitations is small, it may have a
significant effect on the tail of the wounded nucleon distribution at
high multiplicities. These problems will be further discussed in
\sectref{sec:model-final-stat}.

\subsubsection{Impact-parameter profile}
\label{sec:GGTdep}

To investigate further, we need to make assumptions about the
impact-parameter dependence, $T^{(\pp)}(b,\sigma)$ in
\eqref{eq:Tbsigma}.  Strikman \textit{et al.} have suggested a
Gaussian profile on the form
\begin{equation}
  \label{eq:strikgauss}
  T^{(\pp)}(b,\sigma)=\frac{\sigma}{4\pi B}\exp{(-b^2/2B)},
\end{equation}
where $B$ is proportional to $\sigma$. The proportionality factor
could then be fit together with the $\sigma_0$ and $\Omega$ parameters
of \eqref{eq:Psigmatot} to the total and elastic \pp\ cross sections
from \eqref{eq:Tbsigma} and the inclusively wounded cross section in
\eqsref{eq:strikwounded} and \eq{eq:woundedcross}:
\begin{eqnarray}
  \label{eq:strikxcec}
  \sigma\subtot &=& \int d^2b \int d\sigma\,
  P\subtot(\sigma)\,2\, T^{(\pp)}(b,\sigma),\nonumber\\ 
  \sigma\subel &=&  \int d^2b \left|\int d\sigma\, P\subtot(\sigma)\,
    T^{(\pp)}(b,\sigma)\right|^2,\nonumber\\
  \sigma\subwinc  &=& \int d^2b \int d\sigma\,
  P\subtot(\sigma)\left[2\, T^{(\pp)}(b,\sigma) - T^{(\pp)}(b,\sigma)^2\right].
\end{eqnarray}
In this case we find that the wounded cross section distribution can
indeed be written as a simple scaling of the total,
$P\subwinc(\sigma)=P\subtot(\sigma/\lambda\subwinc)$, however, the
same would still not be true for $P\subwabs$.

We also find that for the Gaussian profile, the unitarity constraint,
$T^{(\pp)}(b,\sigma)\le1$, gives a hard limit on
$\sigma\subtot-\sigma\subwinc=\sigma\subel+\sigma\subsdp<\sigma\subtot/4$,
which is not found experimentally.  To proceed we therefore decided to
choose a different form of the $b$ distribution. What is used by
\atlas in \eg\ \cite{Aad:2015zza} is a black disk approximation:
$T^{(\pp)}(b,\sigma)=\Theta(\sqrt{\sigma/2\pi}-b)$. We will instead
use a semi-transparent disk with
\begin{equation}
  T^{(\pp)}(b,\sigma)=T_0\Theta\left(\sqrt{\frac{\sigma}{2\pi T_0}}-b\right),
\label{eq:strikdisk}
\end{equation}
(\cf\ \eqref{eq:gaussianprofile}) where the unitarity constraint gives
us $\sigma\subel+\sigma\subsdp<\sigma\subtot/2$, which can easily
accommodate experimental data.

\subsubsection{Conclusion on the GG formalism}

We conclude that for the \pA\ total cross sections, it is straight
forward to use the GG formalism by Strikman \textit{et al.},
interpreting $\sigma$ as the total cross section averaged over target
states. The distribution $P\subtot(\sigma)$ then describes the
fluctuations in the projectile states. The average and the variance of
$P\subtot$ are given by \eqsref{eq:Psigmatot2} and
\eq{eq:avsigma2}. However, to get the elastic or inclusively wounded
nucleon cross section (including target excitation), we also need to
know the $b$-dependence of $d\sigma(b)/d^2 b \equiv \langle
T^{(\pp)}(b) \rangle_t$ for all values of $\sigma$. If wounded
nucleons are interpreted as absorbed nucleons, we also need to know
$\langle (T^{(\pp)}(b))^2\rangle_t$.  To estimate these quantities in
a way consistent with \eqref{eq:sigmadiff}, we believe it is better to
use a formalism which include individual excitation of both projectile
and target.

\subsection{Consequences of fluctuating \boldmath\pp\ cross section}
Adding fluctuations to the \pp\ cross section dramatically changes the
distributions of the number of wounded nucleons. But since \pp\ data
only offers inclusive and semi-inclusive cross sections to compare
models to, one is given little guidance to why one model works better
than another. Although the \dipsy model is less than perfect in 
reproducing experimental data, it includes those fluctuations in 
the nucleon wave function, which we argue are important when
considering the number of participating nucleons in \pA\ and \AA\ 
collisions. Thus although it only works at high 
energies due to lack of quarks in the proton, the description of 
high-$p_\perp$ particles is poor, and generation of exclusive diffractive
final states is difficult, we believe these deficiencies are less important
when describing the fluctuations.

\subsubsection{Comparison with \dipsy}
\label{sec:comp-with-dipsy}

When comparing the GG results in \eqref{eq:strikxcec} with results from
\dipsy, we first look at the total cross section.  As descussed above, we
interpret the GG fluctuating total cross section $\sigma$ in
\eqref{eq:strikxcec} as describing fluctuations in the projectile, averaged
over target states:
\begin{equation}
	\label{eq:normalxtot}
\sigma\equiv 2 \int d^2 b\, \langle T(b) \rangle_t.
\end{equation}
The parameters in the GG distribution $P\subtot(\sigma)$ can then be tuned to
reproduce the corresponding distribution in \dipsy, which is obtained by
generating a large ensemble of targets for each projectile, and for each target
calculate $T$ at a large number of impact parameters. 

To get the corresponding results for the elastic and the ``wounded'' cross
sections $\sigma\subel=\int d^2 b\, \langle T(b) \rangle_t^2$ and 
$\sigma\subwinc=\int d^2b \left(2 \llangle T\rrangle_t
    - \llangle T\rrangle^2_t\right)$, we have
to make an assumption about the $b$-distribution of the amplitude $\langle
T(b) \rangle_t = T^{(pp)}(b,\sigma)$, appearing in \eqref{eq:strikxcec}. We
here make the simple approximation  in \eqref{eq:strikdisk}, and calculate the
cross section $\sigma\subwinc$ from \eqref{eq:strikxcec}. 
%

Tuning the parameters in $P\subtot(\sigma)$ to the \dipsy results is now 
done fitting the cross sections $\sigma\subtot$, $\sigma\subel$, and
$\sigma\winc$ using a $\chi^2$ fit.
The values obtained with \dipsy are here assigned weights
corresponding to the relative error one would expect from experiment
(taken from the analysis in ref.~\cite{Lipari:2013kta}). The result of
the fit is shown in the first line of table~\ref{tab:dippar}.

\TABULAR{lccr}{ \hline
  & $\Omega$ & $\sigma_0$ & $\lambda$ \\
  \hline
  Original parametrisation & 0.37 & 85.25 & 0.716 \\
  Log-normal parametrisation & 0.25 & 85 & 0.716 \\
  \hline }{\label{tab:dippar} GG parameters values obtained by fit to
  inclusive and semi-inclusive cross sections from \dipsy.}

The result of the fit is compared with the \dipsy results for the
distributions $P\subtot(\sigma)$ and $P\subwinc(\sigma)$ in
\figref{fig:Ptot}. We note here that the $b$-dependence assumed in 
\eqref{eq:strikdisk} implies that the distribution $P(\sigma\subwinc)$ is
given by a scaled $P\subtot$-distribution $P\subwinc(\sigma)\propto
P\subtot(\sigma/\lambda)$, where $\lambda=\sigma\subwinc/\sigma\subtot$.



\FIGURE[t]{
\centering
\begin{minipage}{0.5\linewidth}
  \begin{center}
    (a)
  \end{center}
\end{minipage}\begin{minipage}{0.5\linewidth}
  \begin{center}
    (b)
  \end{center}
\end{minipage}
\includegraphics[width=0.5\linewidth,angle=0]{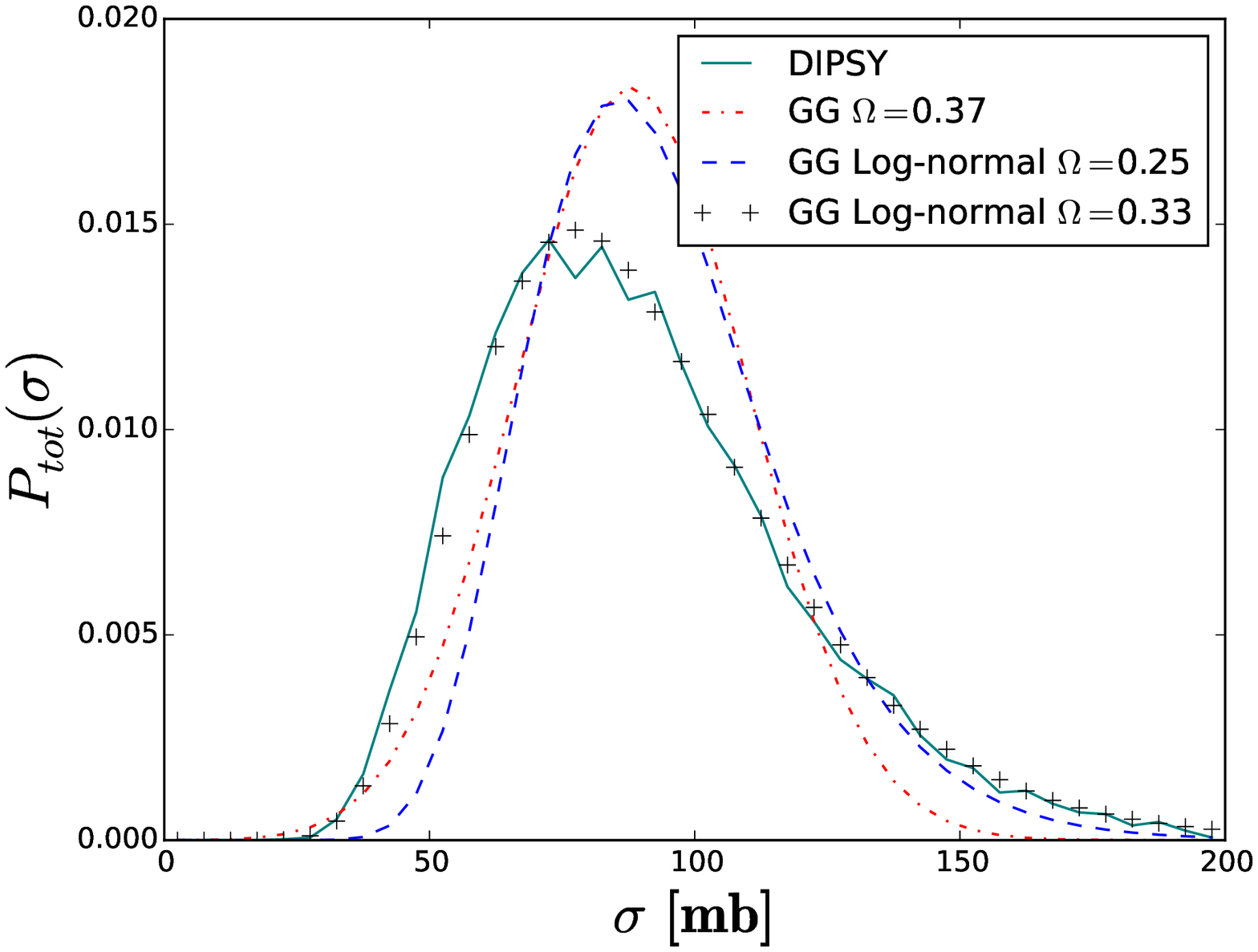}%
\includegraphics[width=0.5\linewidth,angle=0]{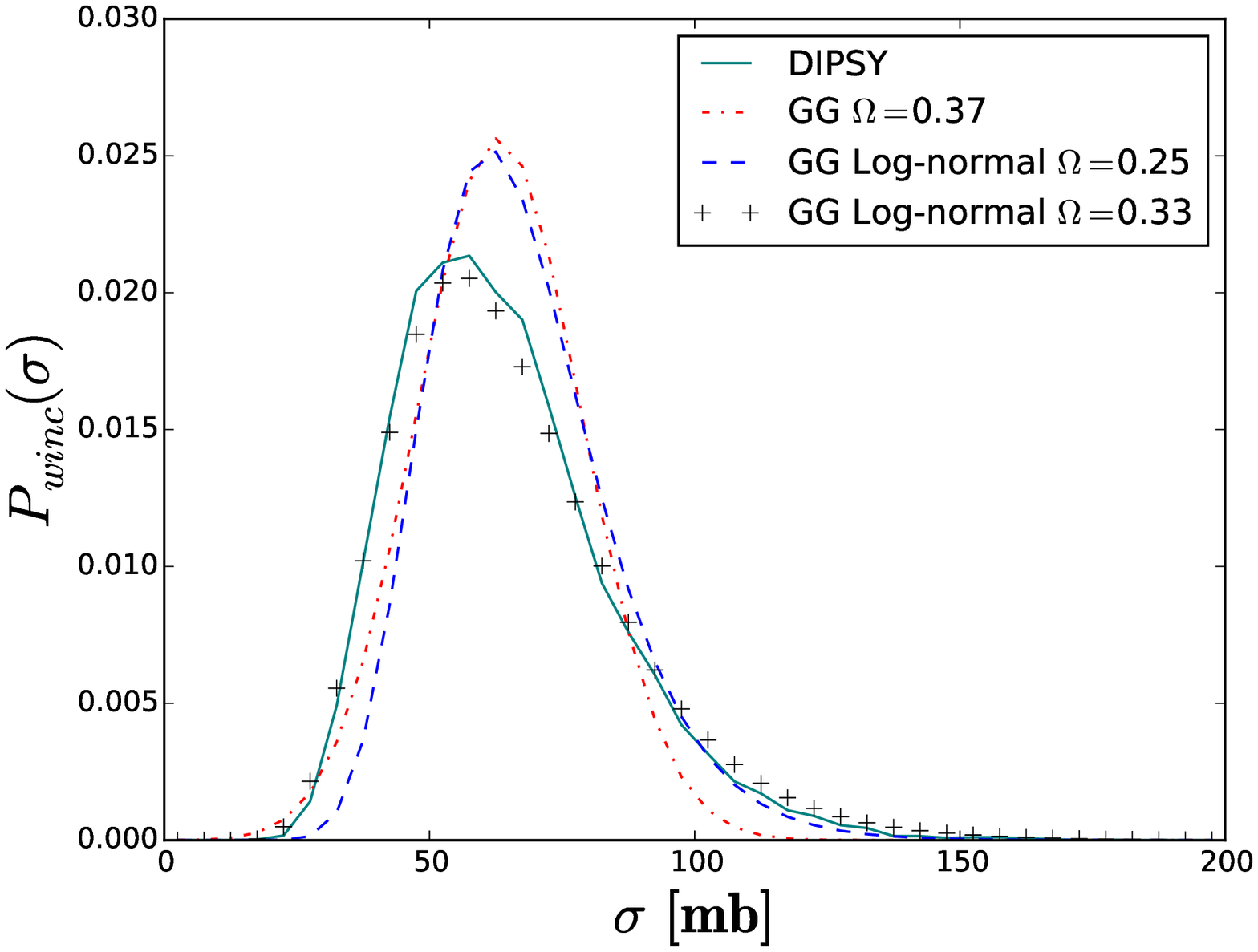}
\caption{\label{fig:Ptot}Fluctuations in the total (a) and inclusively
  wounded (b) cross section by \dipsy and the Glauber--Gribov model
  with different parametrisations of the cross section fluctuations.}
}

It is
clearly seen, that the high-$\sigma$ tails of the \dipsy~distributions
are not reproduced by the functional form for $P\subtot$ in
\eqref{eq:Psigmatot}. Since 
\dipsy~provides a picture of the fluctuations built upon a full
dynamical model, it is reasonable to believe that the shape of the
\dipsy~distributions are closer to reality than
\eqref{eq:Psigmatot}. We therefore try a new parametrisation which
makes it easier to obtain a large high-$\sigma$ tail, namely a
log-normal distribution:
\begin{equation}
  \label{eq:lognormal}
  P_{tot}(\ln \sigma) = \frac{1}{\Omega\sqrt{2\pi}}
  \exp\left(-\,\frac{\ln^2(\sigma/\sigma_0)}{2\Omega^2} \right).
\end{equation}

The fit to the \dipsy cross sections with the log-normal distribution is
also shown in table \ref{tab:dippar}. The corresponding distributions
are shown in figure \ref{fig:Ptot} for two different width parameters, 
labeled $\Omega = 0.25$ and $\Omega = 0.33$. We see that the larger value
matches the \dipsy~distribution perfectly, while the lower value is close to
the GG curve below the maximum but has a higher tail for larger $\sigma$.

\begin{absolutelynopagebreak} 
We note, however, that for technical reasons the diffractive cross section
in \dipsy is calculated demanding a central rapidity gap, restricting the
masses to $M_X^2 \le\sqrt{s}\cdot(1\,\mathrm{GeV})$. This implies that the fluctuations are somewhat underestimated. We therefore
believe that the functional form is quite realistic, while the width 
is underestimated. Results obtained when tuning instead to the experimental
cross sections are presented in the folloing subsection.
\end{absolutelynopagebreak}

\subsubsection{Comparison to data}
\label{sec:gccfcomp}
We now repeat the same procedure as in the previous section, but with
experimental results for the relevant cross sections. 
There is no experimental access to the distributions in cross
section, but the integrated inclusive and semi-inclusive cross
sections are measured, and we here use values from
ref.~\cite{Lipari:2013kta}, extrapolated to $\sqrt{s_{\NN}}=5$~TeV:

\begin{equation}
  \label{eq:expxsec}
  \sigma\subtot = 93.2 \pm 2.3\text{~mb,~}
  \sigma\subel = 23.2 \pm 1.2\text{~mb~and~}
  \sigma\subwinc = 63.0 \pm 1.8\text{~mb}.
\end{equation}
Note that the diffractive cross sections here have been extrapolated into
unmeasured $M_X$ regions to the full $0<M_X<\sqrt{S}$ interval. As mentioned
above this was not done in for the DIPSY diffractive cross sections in
\sectref{sec:comp-with-dipsy}, where by construction a rapidity gap is
required at mid rapidity. Hence we expect that the fluctuations for DIPSY are
underestimated as compared to data. The parameter values obtained by
minimising the $\chi^2$ are listed in table~\ref{tab:datpar}. 

\TABULAR{lccr}{
  \hline
  & $\Omega$ & $\sigma_0$ & $\lambda$ \\
  \hline
  Original parametrisation & 0.82 & 77.75 & 0.677 \\
  Log-normal parametrisation & 0.43 & 85 & 0.677 \\
  \hline }{\label{tab:datpar} GG parameter values obtained by fit to inclusive and semi-inclusive cross-section from data.}

In \figref{fig:datafitSigma} we compare the fits of the two
parametrisations of $P\subwinc(\sigma)$ shown above.  We
see that the new parametrisation provides larger fluctuations in the
high-$\sigma$ tail, as expected. It should be noted that both fits
reproduce the experimental cross sections well within the experimental
errors. For comparison we also show log-normal distribution 
$P\subwinc(\sigma)$, when it's width and mean are fitted to \dipsy~by eye  
(denoted $\Omega=0.33$), which is significantly more narrow.

\FIGURE[t]{
  \centering
  \includegraphics[width=0.6\linewidth,angle=0]{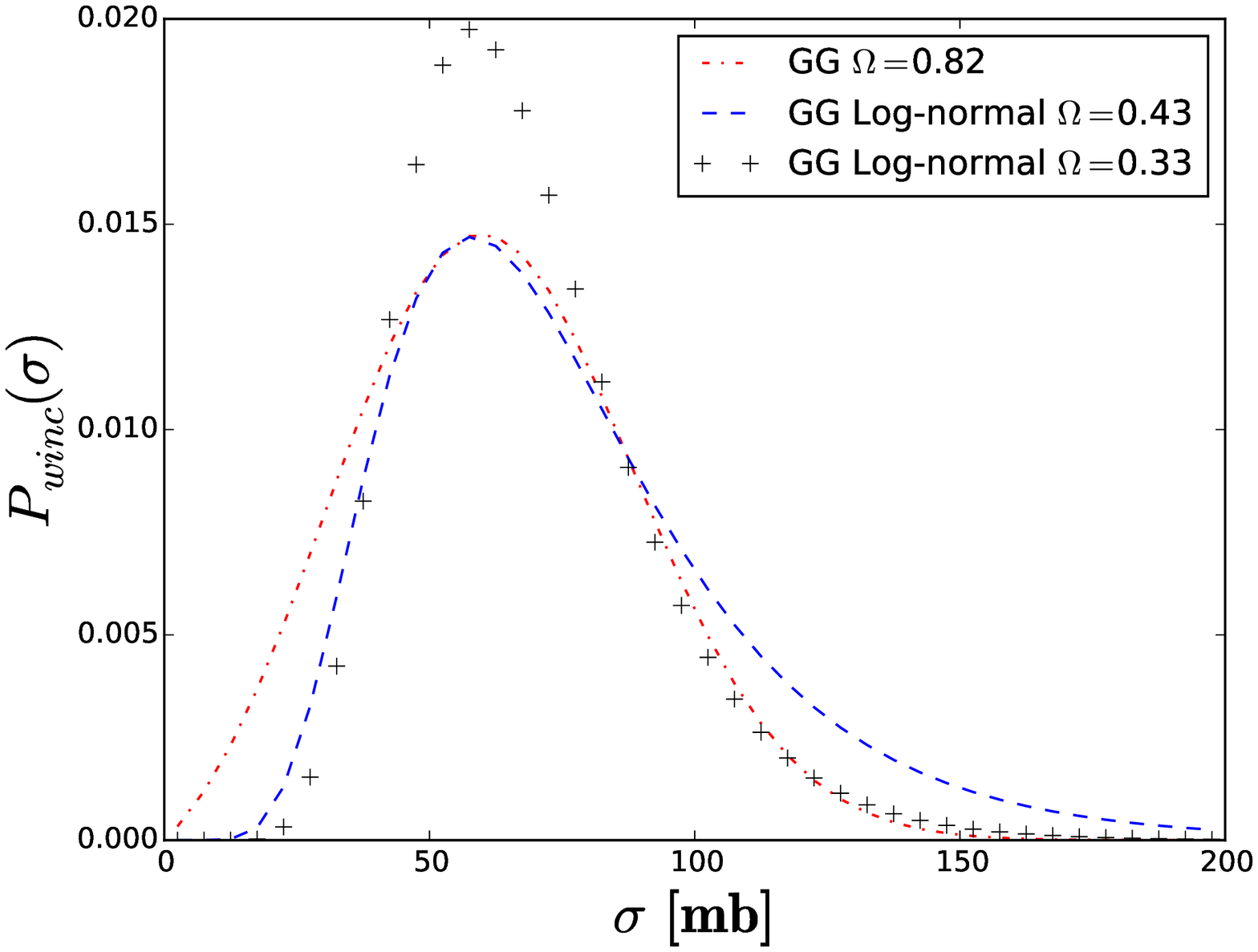}
  \caption{Fluctuations in the inclusively wounded cross section by
    the Glauber--Gribov model with two parametrisations of the cross
    section fluctuations, fitted to data.}
\label{fig:datafitSigma}}

We suspect that while the log-normal parametrisation probably gives a
more realistic description  of the high-$\sigma$ fluctuations in the GG
formalism, it is far from the whole story. The GG results presented here are
obtained assuming that all fluctuations are ascribed to fluctuations in
projectile \textit{size}, as described in 
\sectref{sec:GGTdep}. In \dipsy, however, the cross section
fluctuations arise from a combination of fluctuations in size and
fluctuations in gluon density. We believe that updating the profile
functions from simple disks or Gaussian distributions to more
realistic ones, could provide a better handle on the parametrisations
of the cross section from \pp\ data, this will be investigated in a
future publication.  So far we have described a prescription which
seems to both catch the necessary physics to calculate the inclusive
wounded cross section, with all parameters being obtainable from
\pp~data. We will now apply this to \pA~collisions.

\subsection{Distributions of wounded nucleons}
\label{sec:woundeddist}

Using the considerations about fluctuations in the wounded cross
section, we will now turn to generation of distributions of wounded
nucleons. Normally, in inelastic, non-diffractive \pA\ collisions, the
number of wounded nucleons is always one plus the number of inelastic,
non-diffractive \NN\ interactions. In the following we will make the
distinction between diffractively and absorptively wounded
nucleons. In order to avoid situations where the projectile should
sometimes be counted twice as a wounded nucleon, we will solely talk
about the number of wounded nucleons in the target, which we denote
$N^t\subw$. We note also that since the number of sub-collisions and the
number of wounded nucleons are trivially connected, the question
whether a specific observable scales better with wounded nucleons or
with \NN\ sub-collisions, is much more relevant for nucleus--nucleus
collisions.

\subsubsection{Inclusively wounded nucleons}

We will describe the nucleus' transverse structure using a
Woods--Saxon distribution in the GLISSANDO parameterisation
\cite{Broniowski:2007nz,Rybczynski:2013yba}, where the density is
given by:
\begin{equation}
  \rho(r) = \frac{\rho_0 (1 + wr^2/R^2)}{1+\exp((r-R)/a)},
\label{eq:woodsaxon1}
\end{equation}
where $R$ is the nuclear radius, $a$ is ``skin width'', and $\rho_{0}$
is the central density. The parameter $w$ describes a possible
non-constant density, but is zero for lead. The nucleons are generated
with a hard core, which thus introduces short range correlations among
the nucleons. As shown by Rybczynski and
Broniowski~\cite{Rybczynski:2010ad}, the correct two-particle
correlation can be obtained if the nucleons are generated with a
minimum distance equal to $2r\sub{core}$. Using $r\sub{core}=0.45$~fm
and a skin width of $a=0.459$, the radius of the Lead nucleus becomes
$R^{\Pb}=6.406$ according to the parameterisation in
\cite{Rybczynski:2013yba}.

For each nucleus state we generate a random impact parameter wrt.\ the
projectile proton and proceed to determine which nucleons will be
wounded, following the previously outlined models.

In \figref{fig:npartFluct}, we show the distribution in the number of
inclusively wounded nucleons (using $\sigma\subwinc\sup{\pp} =
63.0$~mb) for: a black disk model without any \pp\ cross section
fluctuations; GG with parameters fitted to data in
\sectref{sec:gccfcomp}; GG with $P_{tot}(\sigma)$ given by
\eqref{eq:lognormal}, also fitted to data; and the new simplified
model outlined in \sectref{sec:simple-models} (here called \newdisk)
Fitting the latter to the cross sections in \eqref{eq:expxsec} as well
as to the double diffractive cross section $\sigma_{DD} = 3.2$ mb, we
obtain the parameters listed in table~\ref{tab:greypar}.

\TABULAR{lccr}{ 
  \hline
  $r_1$ & $r_2$ & $\alpha$ & $c$ \\
  \hline
  0.15 fm & 1.07 fm & 0.97 & 0.42 \\
  \hline 
  }{\label{tab:greypar} Table of parameters of the \newdisk
  model fitted to \pp\ data.}

\FIGURE[t]{
\centering
\includegraphics[width=0.6\linewidth,angle=0]{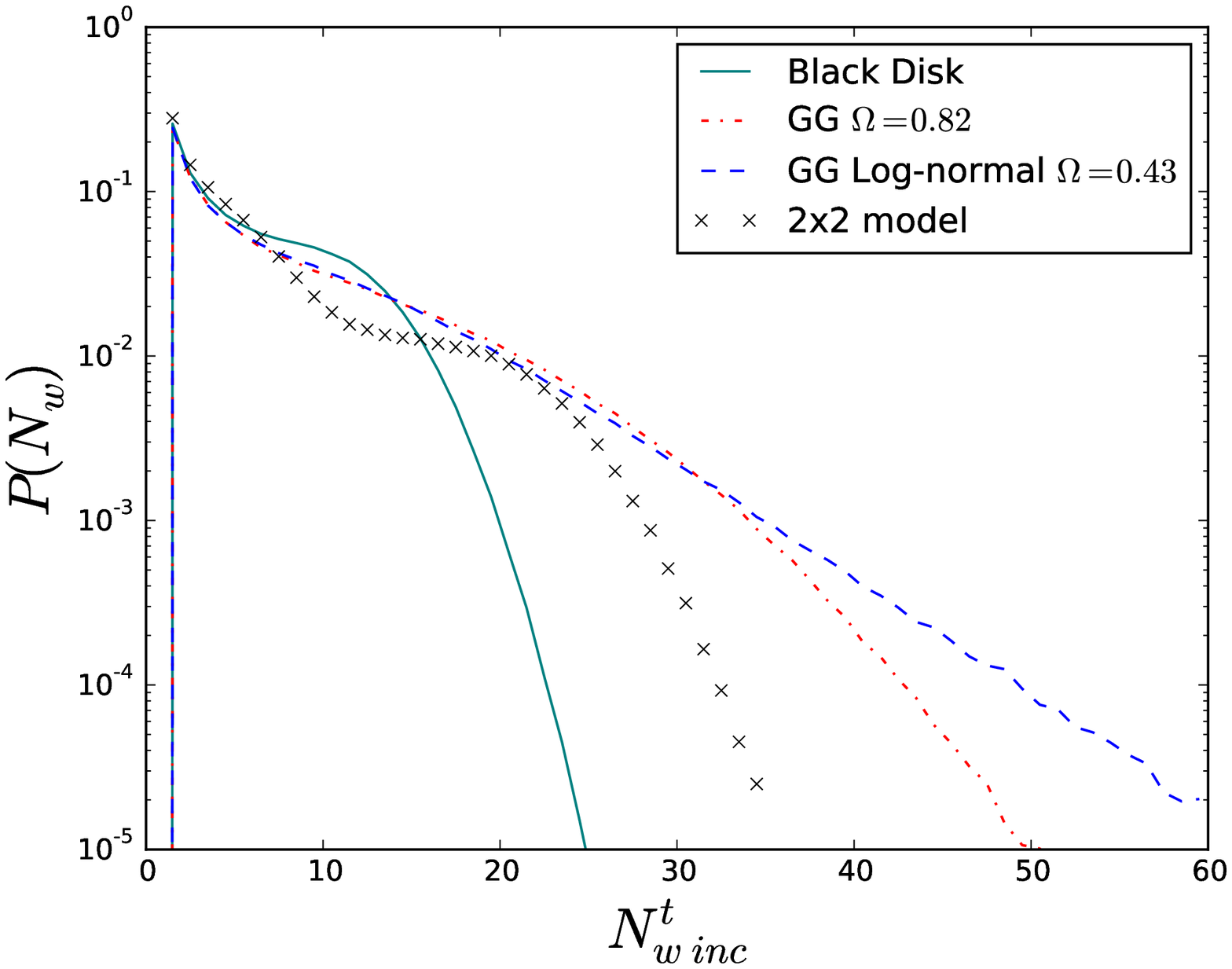}
\caption{Distribution in the number of inclusively wounded nucleons,
  $N^t\subwinc$, in \pPb\ events at $\sqrt{s_{NN}}=5$~TeV, for a
  Glauber black disk, the GG model with two parametrisations of
  $P_{tot}(\sigma)$ and the \newdisk model. All models have been
  fitted to reproduce relevant measured (semi-) inclusive cross
  sections.}
\label{fig:npartFluct}
}

Looking at the individual distributions in \figref{fig:npartFluct}, we
see that all three inclusions of additional fluctuations in the cross
section, significantly increases the tail of the distribution compared
to the black disk. The \newdisk model has fewer fluctuations to very large
$N\subwinc^t$ numbers, and the
dip in the distribution around $N^t\subwinc = 10$, also indicates that
the fluctuations are too crude. The difference between GG with the
original parametrisation and the log-normal distribution is visible in
the tail above $N^t\subwinc \approx 35$, as expected. One would
therefore expect only an effect in the central events.

\subsubsection{Distinguishing between absorptively and diffractively wounded
  nucleons} 
\label{sec:distinguish}
In our interpretation of the GG model in \sectref{sec:strikdiff}, it
can be used to calculate the sum of absorptively and diffractively
wounded nucleons. In the Monte Carlo one would, however, like to have
an impact parameter dependent recipe for each sub-collision to decide
whether or not a target nucleon is diffractively or absorptively
wounded, when hit by a projectile in a definite state $p$. This amounts to
calculating the ratio of the absorptive to 
the inclusively wounded cross sections for a given sub-collision, and
compare it to a random number

\begin{equation}
  \label{eq:mcdiff}
  \frac{P_{\wabs,p}}{P_{\winc,p}} =
  \frac{2\llangle T_{p,t}(b)\rrangle_t - \llangle T_{p,t}(b)\rrangle_t^2}{2\llangle T_{p,t}(b)\rrangle_t - \llangle T^2_{p,t}(b)\rrangle_t}.
\end{equation}

For the \newdisk model this is done easily, as the above ratio reduces
to:

\begin{equation}
  \frac{P_{\wabs,p}}{P_{\winc,p}} =
  	\frac{2-\alpha}{2-\alpha\llangle T^2_{p,t}(b)\rrangle_t}.
\end{equation}
The GG model on the other hand, implies averaging over target nucleon
states, and provides thus no distinction. Instead we follow the
\newdisk model to calculate the the conditional probability to be
diffractively wounded, if a nucleon is already inclusively wounded in
the GG model. This is:

\begin{equation}
	P(\mrm{diff} | \winc) = \Theta\left(\sqrt{\sigma_{GG}/\pi} - (r_1 -
        r_2) - b\right)\frac{2-\alpha}{2-\alpha c}, 
\end{equation}
where the first term is a requirement that the two nucleons are
separated by an amount such that a fluctuation in size is necessary to
be wounded.  In \figref{fig:npartAbs} we show distributions of
$N^t\subwabs$ for the \newdisk model and for the corrected GG model,
using both parametrisations of $P(\sigma)$.  \FIGURE[t]{ \centering
\includegraphics[width=0.6\linewidth,angle=0]{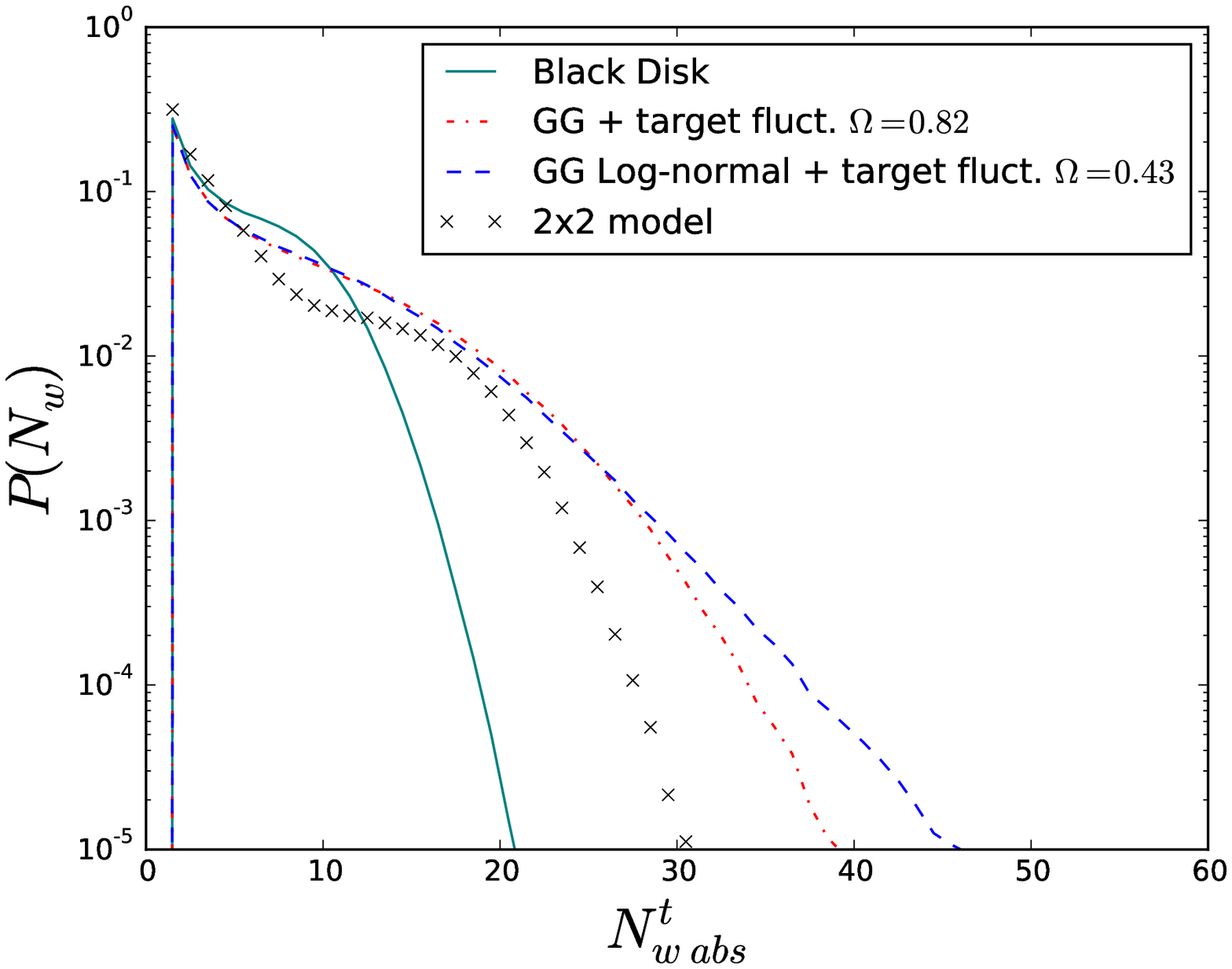}
\caption{Distribution in the number of absorptively wounded nucleons,
  $N^t\subwabs$, in \pPb\ events at $\sqrt{s_{NN}}=5$~TeV, for a
  Glauber black disk, the GG model with two parametrisation of
  $P(\sigma)$, corrected using the \newdisk model, along with the
  \newdisk model itself.  All models have been fitted to reproduce
  relevant measured (semi-) inclusive cross sections.}
\label{fig:npartAbs}
}

\section{Modelling final states in \boldmath\pA\ collisions}
\label{sec:model-final-stat}
In this section we will take the knowledge about distributions of
wounded nucleons and investigate the consequences for final states in
\pA\ collisions. We will discuss a few views on modelling particle
production in such collisions, all assuming that a full final state of
a \pA\ collision can be adequately modelled by stacking \pp\ events on
top of each other, here modelled using \pytppp.  Following the
introduction of the models, we will compare to data, both multiplicity
as function of centrality, and inclusive $p_\perp$ spectra. Finally we
will give an estimate of the theoretical uncertainties present at this
early stage of the model.

\subsection{Generating final states with \pytppp}
\label{sec:finalstates}
The general methodology for generating final states, which will be
pursued here, will have the following ingredients:
\begin{itemize}
\item For each collision, a Glauber calculation is performed as
  outlined in \sectref{sec:woundeddist}, setting up the nuclear
  geometry.
\item The total number of inclusively wounded target nucleons is calculated,
  as well as the number of absorptively wounded targets, if the two
  differ in the considered approach.
\item Sub-collisions are generated as \pp\ collisions, according to
  two separate approaches, which will be outlined in the following.
\item Each sub-collision is treated separately in terms of colour
  reconnection and hadronisation. Efforts to include cross talk
  between sub-collisions will be the subject of a future publication.
\end{itemize}
Cross talk between sub-collisions is, however, included in one respect
by accounting for energy-momentum conservation in all approaches. As
before, we will concentrate on \pPb\ collisions at
$\sqrt{s_{\NN}}=5$~TeV. The methodology is, however, not limited to
this, and generalisation to \AA\ collisions will be the subject of a
future publication.

\subsection{Wounded nucleons and multi-parton interactions}

In ref. \cite{Bialas:1976ed} Bia\l as \textit{et al.} noticed that the central particle
density in \pA\ collisions scales approximately with the number of "wounded" or
"participating" nucleons, $dN^{(\pA)}/d\eta \approx
(N_{wounded}/2)\,dN^{(\pp)}/d\eta$. The projectile proton was here included as one
of the wounded nucleons, and the distribution in rapidity could be described
if each wounded target nucleon gives a contribution proportional to $(\eta -
\eta_t)/(\eta_p-\eta_t)$ where $\eta_{t,p}$ are the rapidities of the target
and projectile respectively. The wounded projectile proton gives a similar
contribution with $p$ exchanged for $t$.

The wounded nucleon model worked well for minimum bias events and low
$p_\perp$ particles, while high $p_\perp$ particles scale better with the
number of 
$NN$ collisions, which can be understood if the high-$p_\perp$ particles
originate from independent partonic subcollisions. (See
\eg\ ref.~\cite{Bialas:9999aa}.) A model with this feature,
called G-Pythia, has been used in analyses by \alice
\cite{Adam:2014qja}.

These results can be given a heuristic interpretation in terms
of the Landau-Pomeranchuk formation time. The formation time  
for a hadron is, in a frame where $p_L=0$:
\begin{equation}
	\tau \geq \frac{1}{\sqrt{p_\perp^2 + m^2}}.
\end{equation}
This implies that a produced pion will resolve the
nucleus  
at a length scale given roughly by $1/p_\perp$.
For $p_\perp < 1$ GeV the resolution scale is larger than that of
the individual nuclei, while for $p_\perp$ larger than $\sim 1$ GeV,
constituents of individual nucleons can be resolved. 

Below we will compare two models, generated with the help of \pythia8. The
model denoted "Absorptive" is similar to G-Pythia. Here each $NN$ subcollision
is treated as a \pp\ collision\footnote{Note that the G-Pythia approach really
  uses a black 
  disk Glauber calculation with $\sigma_{\NN} = \sigma\subineltot=
  \sigma\subabs+\sigma\subsdp+\sigma\subsdt+\sigma\subdd$, and lets the
  collisions be a mixture of the four corresponding processes, while
  we we consider only absorptive collisions, as we believe this is
  more in line with the original model by Glauber. There is, however,
  no notable difference for observables in the near central
  rapidity range, taken with a minimum bias trigger.}.
The second model, explained in the
next subsection, is called FritiofP8 and is more similar to the wounded nucleon
model. We note that the models should not be compared on equal footing. From the above arguments, the "Absorptive"
model is expected to describe the high-$p_\perp$ part of the spectrum better, while FritiofP8, being similar to the wounded nucleon model,
is expected to describe the low-$p_\perp$ part, and thus also the total multiplicity, best.

Technically, the subcollisions are in both models generated with \pytppp. This
means that for each sub-collision, 
multiple partonic interactions are created in decreasing order of $p_\perp$
with the probability: 
\begin{equation}
	\label{eq:pythiaMpi}
	\frac{d\mathcal{P}}{dp_{\perp i}} =
        \frac{1}{\sigma\subabs}\frac{d\sigma_{2\rightarrow 2}}{dp_{\perp i}}
        \exp\left[-\int_{p_{\perp i}}^{p_\perp i-1}
          \frac{1}{\sigma\subabs}
          \frac{d\sigma_{2\rightarrow 2}}{dp_\perp'}dp_\perp' \right],
\end{equation}
starting from a maximum scale related to the impact parameter of the
sub-collision. The cross section is obtained by treating
everything as perturbative QCD $2\rightarrow 2$ scatterings, but since
the cross section diverges at low $p_\perp$, it is regulated at low
$p_\perp$ using:
\begin{equation}
  \label{eq:mpixsec}
  \frac{d\sigma_{2\rightarrow 2}}{dp_\perp^2} \propto
  \frac{\alpha^2_s(p^2_\perp)}{p^4_\perp} \rightarrow
  \frac{\alpha^2_s(p^2_\perp + p^2_{\perp 0})}{(p^2_\perp + p^2_{\perp 0})^2}.
\end{equation} 
Here $p_{\perp 0}$ is a tunable parameter.

Aside from momentum conservation, \pytppp also rescales the PDF every
time a quark has been used in an MPI. When using this MPI model for
generating \pA\ collisions we maintain momentum conservation, but do
not maintain the rescaling of the PDF between separate \NN\
collisions.

\subsection{The revived Fritiof model}
\label{sec:fritiof}
A very different approach was used in the Fritiof model
\cite{Andersson:1986gw}. Where the \pytppp\ MPI model assumes
everything can be described by perturbative scatterings, Fritiof
imposed a soft model for everything, specifically limiting it's range
of validity to low-$p_\perp$ processes.\footnote{A motivation for the
  development of the Fritiof model, was to get a realistic
  extrapolation from \pp\ collisions to collisions with nuclei. This
  could then form a background in searches for possible collective
  effects. Unfortunately it worked too well (at the energies available
  in the eighties), basically leaving no evidence for plasma
  formation.}

In the Fritiof model it is assumed that a soft min-bias interaction
causes a momentum exchange, which in light-cone variables has the form
\begin{equation}
    \label{eq:fritiof}
    P(Q_+,Q_-) \propto \frac{dQ_+}{Q_+}\frac{dQ_-}{Q_-}.
\end{equation}
This produces two excited states assumed to decay like strings
stretching the rapidity range between the initial beam rapidities and
a point distributed evenly within the kinematically allowed region The
result is approximately reproducing the original wounded nucleon model
\cite{Bialas:1976ed}, but it is in the Fritiof model also assumed that
a secondary encounter with another nucleon will increase the
excitation, thus leading to a logarithmic scale breaking.

In ref. \cite{Andersson:1992iq} it was suggested to extend the Fritiof
model to include the possibility for a hard scattering and associated
bremsstrahlung when the energy is high enough. At LHC collision
energies, the necessity for including the possibility for at least one
such interaction is apparent.

\FIGURE[t]{  \centering
\scalebox{0.8}{\mbox{ 
  \begin{picture}(300,300)(0,0)
    \GOval(0,150)(10,7)(0){1}
    \Text(-12,150)[]{p}
    \GOval(300,50)(10,7)(0){1}
    \Text(318,50)[]{$\nu_1$}
    \GOval(300,100)(10,7)(0){1}
    \Text(318,100)[]{$\nu_2$}
    \GOval(300,150)(10,7)(0){1}
    \Text(318,150)[]{$\nu_3$}
    \GOval(300,200)(10,7)(0){1}
    \Text(318,200)[]{$\nu_4$}
    \GOval(300,250)(10,7)(0){1}
    \Text(318,250)[]{$\nu_5$}

    \SetColor{Red}
    \Gluon(0,150)(80,150){4}{8}
    \Gluon(80,150)(160,165){4}{8}
    \Gluon(160,165)(220,200){4}{6}
    \Gluon(220,200)(300,200){4}{8}

    \SetColor{Gray}
    \Gluon(240,250)(300,250){4}{6}
    \ZigZag(240,250)(220,200){4}{5}
    \GOval(240,250)(7,5)(0){1}
    \ColText(240,225)[]{$I\!\!P$}

    \SetColor{Blue}
    \Gluon(160,165)(300,150){4}{14}

    \SetColor{Green}
    \Gluon(80,150)(220,95){4}{14}
    \Gluon(220,95)(300,50){4}{8}
    
    \SetColor{Orange}
    \Gluon(220,95)(300,100){4}{8}

    \SetColor{Black}

  \end{picture}}}

\caption{Cartoon in rapidity--impact-parameter space, showing the
  evolution of exchanged gluons between a projectile proton and a
  number of wounded nucleons in the target nucleus. Nucleons
  $\nu_1,\ldots,\nu_4$ are wounded absorptively, while $\nu_5$ is
  wounded diffractively. $\nu_4$ is considered to be the primary
  wounded nucleon.}
  \label{fig:cartoon}
}

In \figref{fig:cartoon} we show a schematic picture of a projectile
proton wounding a number of nucleons. The picture is strongly
oversimplified, showing only the main gluon propagators, \ie\ no
initial/final-state radiation (ISR/FSR) or multi-parton
interactions (MPI) in the individual sub-collision. Nucleons $\nu_1,
\ldots \nu_4$ are wounded absorptively with $\nu_4$ being the hardest
or "primary" wounded nucleon, which contributes to the hadronic
multiplicity the full rapidity region. The other absorptively wounded
nucleons, $\nu_1 \ldots \nu_3$, contributes only to the parts of the
rapidity range in the nucleus direction. As indicated by the exchanged
Pomeron, $I\!\!P$, $\nu_5$ is only diffractively excited, and will
also only contribute in the nucleus direction.

\FIGURE[t]{  \centering
\scalebox{0.8}{\mbox{ 
  \begin{picture}(300,300)(0,0)
    \GOval(50,220)(7,7)(0){1}
    \GOval(10,80)(7,7)(0){1}
    \GOval(90,80)(7,7)(0){1}
    \def\axowidth{1.5 }    
    \Line(10,220)(43,220)
    \Line(57,220)(90,220)
    \Line(-20,80)(3,80)
    \Line(17,80)(45,70)
    \Line(55,70)(83,80)
    \Line(97,80)(120,80)
    \def\axowidth{0.5 }
    \ZigZag(50,213)(50,150){6}{6}
    \ZigZag(50,150)(10,87){6}{6}
    \ZigZag(50,150)(90,87){6}{6}
    \SetColor{Red}
    \DashLine(50,250)(50,50){4}
    \SetColor{Black}

    \GOval(250,220)(7,7)(0){1}
    \GOval(210,88)(7,7)(0){1}
    \GOval(290,72)(7,7)(0){1}
    \def\axowidth{1.5 }    
    \Line(210,220)(243,220)
    \Line(257,220)(290,220)
    \Line(180,88)(203,88)
    \Line(217,88)(273,88)
    \Line(291,88)(320,88)
    \Line(180,72)(195,72)
    \Line(180,72)(283,72)
    \Line(297,72)(320,72)
    \def\axowidth{0.5 }
    \ZigZag(250,213)(250,150){6}{6}
    \ZigZag(250,150)(212,95){6}{6}
    \ZigZag(250,150)(287,79){6}{6}
    \SetColor{Red}
    \DashLine(250,250)(250,150){4}
    \DashLine(250,150)(202,76){4}
    \DashLine(198,70)(191,60){4}
    \DashLine(250,150)(300,50){4}
    \SetColor{Black}
    \Text(50,40)[]{(a)}
    \Text(250,40)[]{(b)}

  \end{picture}}}

\caption{Pomeron diagrams with cuts indicated for (a) single
  diffractive excitation in proton--proton and (b) doubly absorptive
  proton--deuteron scattering.}
\label{fig:cutpom}
}

Thinking in terms of cut Pomeron diagrams \`a la AGK
\cite{Abramovsky:1973fm} we show in \figref{fig:cutpom} the
similarity between the diagram describing diffractive excitation in
proton--proton scattering and a fully absorptive proton--deuteron
scattering. It is not far fetched to assume that
the triple-Pomeron vertex in both cases are distributed in
approximately the same way in rapidity, \ie, that the gap size in the
single diffractive excitation in \pp\ would be distributed in the same
way as the the size of the region of rapidity populated by hadrons
from both wounded nucleons in a \pd\ collision.

As discussed in \sectref{sec:wounded} the distribution in diffractive
masses indicates a fairly flat distribution in rapidity of the
triple-Pomeron vertex as $\epsilon$ is close to zero\footnote{The
  default in \pytppp is actually to have $\epsilon=0$ for high-mass
  diffraction, which corresponds to the distribution used in Fritiof
  in \eqref{eq:fritiof}.}. We will therefore assume as a first
approximation, that the secondary absorptive collisions in a \pA\
collision can be approximated by single diffractive collisions.

Treating secondary absorptive collisions as single diffractive
excitation has an additional added benefit. In the \pytppp
implementation, one can model high mass soft excitation using a
perturbative approach where the exited proton can undergo multiple
partonic interactions, as in \eqref{eq:pythiaMpi}, and ISR is included. It
is thus possible to treat absorptively wounded nucleons differently,
depending on whether the mass of the excited system is larger or smaller than a
pre-set threshold mass scale.

Finally, in \figref{fig:cartoon}, we also have $\nu_5$, which is an
standard diffractively excited nucleon, and will be modelled as such.

\subsection{Comparison to data}
We now compare the two methods for particle production, which were
introduced above. Stacking absorptive events on top of each other is
labelled ``\ourabsorptive'', we use a black disk Glauber model with
$\sigma_{abs} = 67.9$ mb to calculate the number of absorptive
sub-collisions event by event. The model including both diffractive
excitation and the Fritiof-inspired absorptive sub-collisions is
labelled ``\newfrit''. To calculate the amount of wounded nucleons we
use the modified GG model with cross section fluctuations described by
the log-normal distribution in \eqref{eq:lognormal} and including the
modifications introduced in \sectref{sec:GGTdep}, as well as
distinguishing between absorptive and diffractive events using the
\newdisk modification, introduced in \sectref{sec:distinguish}. All
parameters are fitted to \pp~data.

\subsubsection{Centrality estimation and multiplicity}

The primary observable we wish to discuss, is the charged particle
pseudo-rapidity distribution at different centralities, as measured by
\atlas \cite{Aad:2015zza}. When comparing Monte Carlo predictions to
data in \pp, the work flow has matured greatly over the past years,
with the advent of automated frameworks for performing such tasks,
such as Rivet \cite{Buckley:2010ar}. In this framework, equal
treatment of theory and unfolded data is ensured by publishing
measurements along with an implementation of the analysis. This is not
yet tradition in the heavy ion community, and the data comparisons
shown here, is the result of our own Rivet implementation on the
analysis, based on the paper, with data obtained from HepData
\cite{Buckley:2006np}.

In the experimental analysis by \atlas, event centrality is calculated
by taking fractiles of the distribution in $\sum E_\perp$ of charged
particles in the interval\footnote{Notice that our definition of
  $\eta$ is opposite to the one used in ref. \cite{Aad:2015zza}, but
  follows the HepMC published data.} $3.1 < \eta < 4.9$. For this
particular observable, unfolded data has not been published, but we
will still compare theoretical curves for the two previously outlined
particle production models. In \figref{fig:sumet} we show the model stacking
$N^t\subwabs$ absorptive events spanning the whole rapidity region
(denoted ``\ourabsorptive'') reaches a much higher $\sum E_\perp$ than
the Fritiof inspired model (denoted ``\newfrit'') With one absorptive event
spanning the 
whole rapidity region, $N^t\subwabs-1$ absorptive events modelled as
diffractive excitation, and $N^t\subwinc-N^t\subwabs$ events from
diffractive excitation. We note that the ``\newfrit'' results agree almost
perfectly with the data from \textsc{Atlas}, while the \ourabsorptive model
reaches significantly higher $\sum E_\perp$ values.

\FIGURE[t]{
\centering
\includegraphics[width=0.6\linewidth,angle=0]{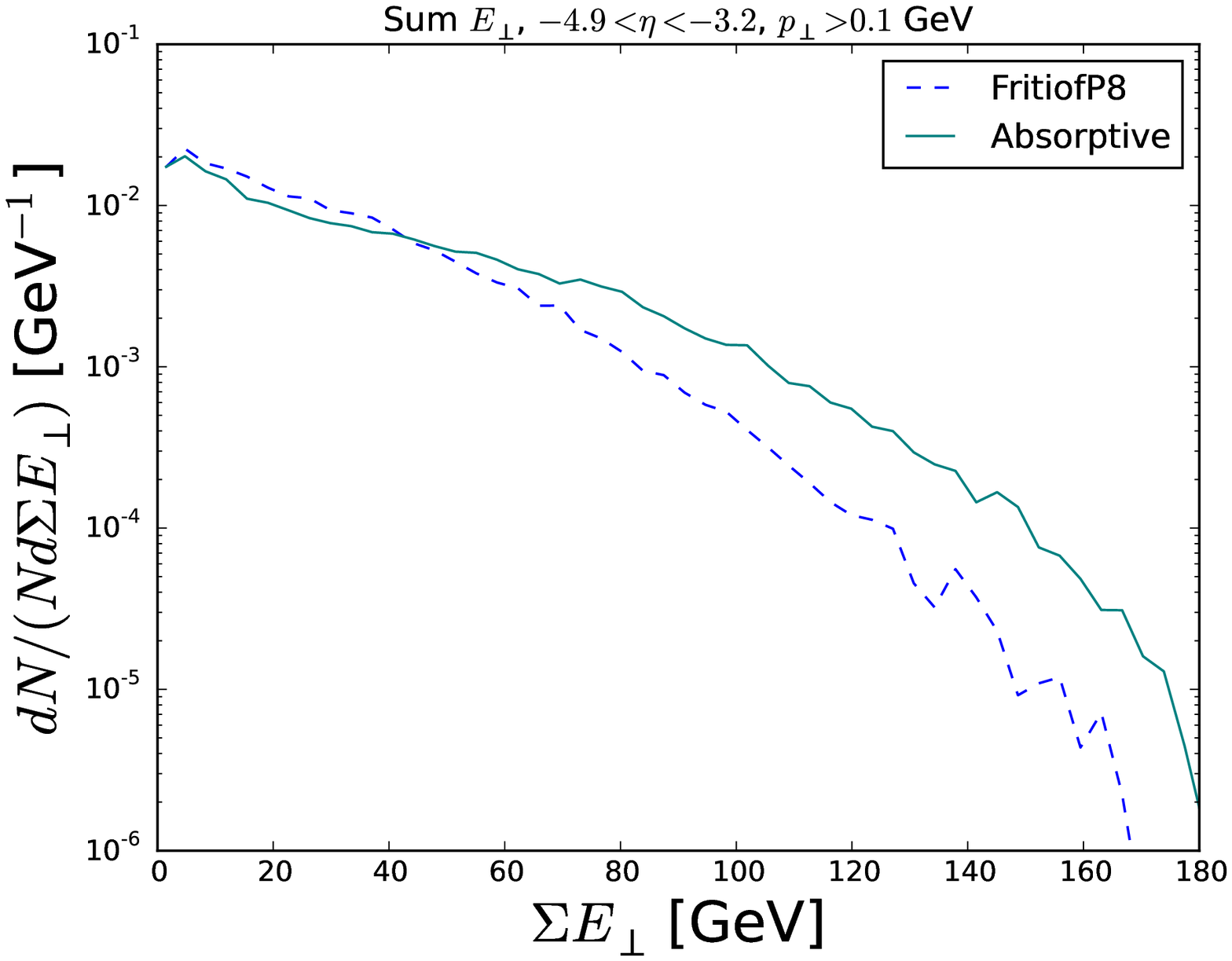}
\caption{Distribution in $\sum E_\perp$ for a sum of full absorptive
  events and the new \newfrit model, for \pPb\ collisions at $\sqrt{s_{NN}}=5$
  TeV.} 
\label{fig:sumet}
}

In \figref{fig:multPeriph} we show pseudorapidity distributions for different
centralities, where we have used the same cuts as \atlas, but
reconstructed fractiles from our own distribution. 
\footnote{Further centralities are shown on
 \texttt{http://home.thep.lu.se/DIPSY/FritiofP8}, but omitted here
 for brevity.} 
\FIGURE[t]{
\centering
\begin{minipage}{0.5\linewidth}
  \begin{center}
    (a)
  \end{center}
\end{minipage}\begin{minipage}{0.5\linewidth}
  \begin{center}
    (b)
  \end{center}
\end{minipage}
\includegraphics[width=0.5\linewidth,angle=0]{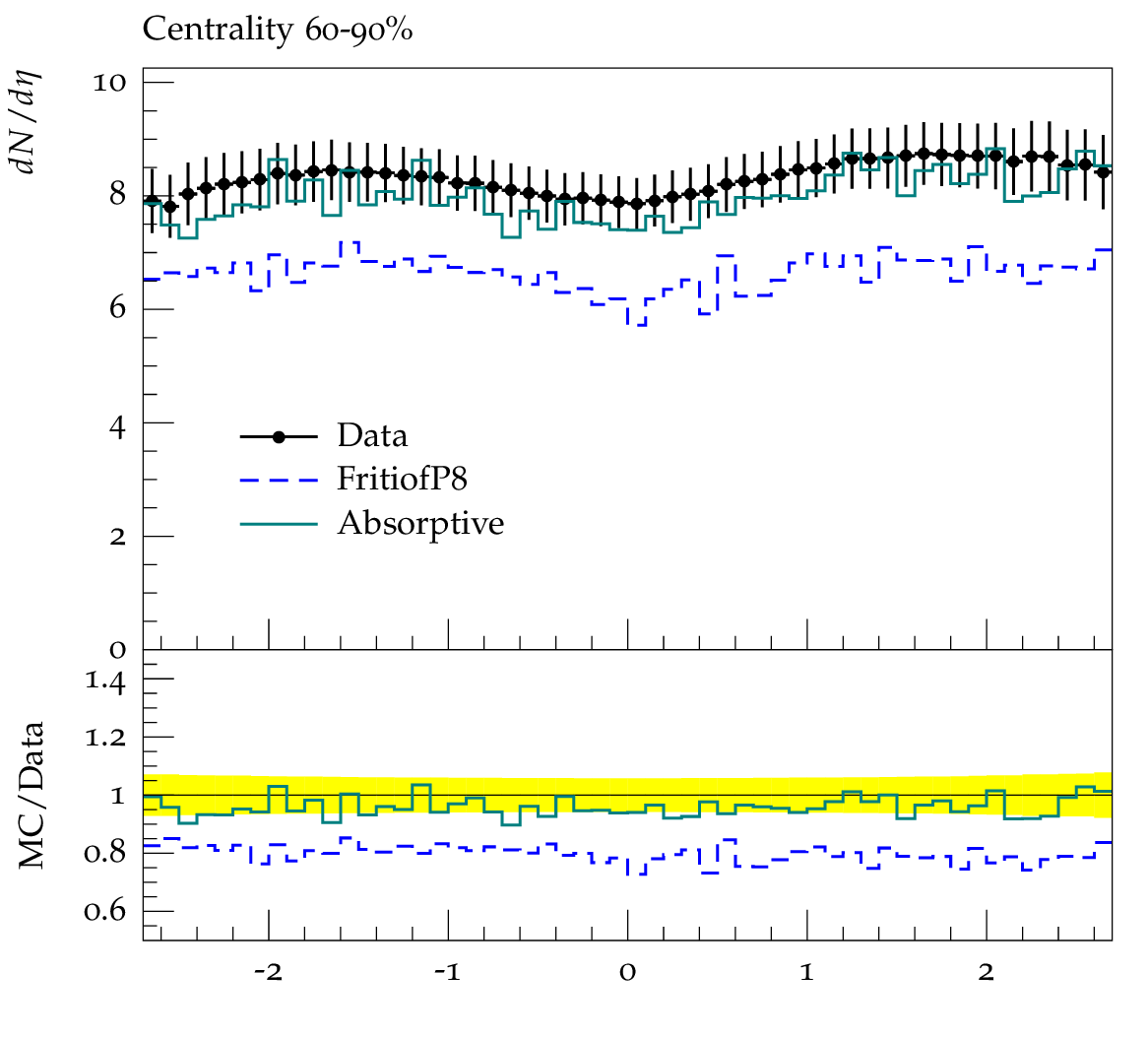}%
\includegraphics[width=0.5\linewidth,angle=0]{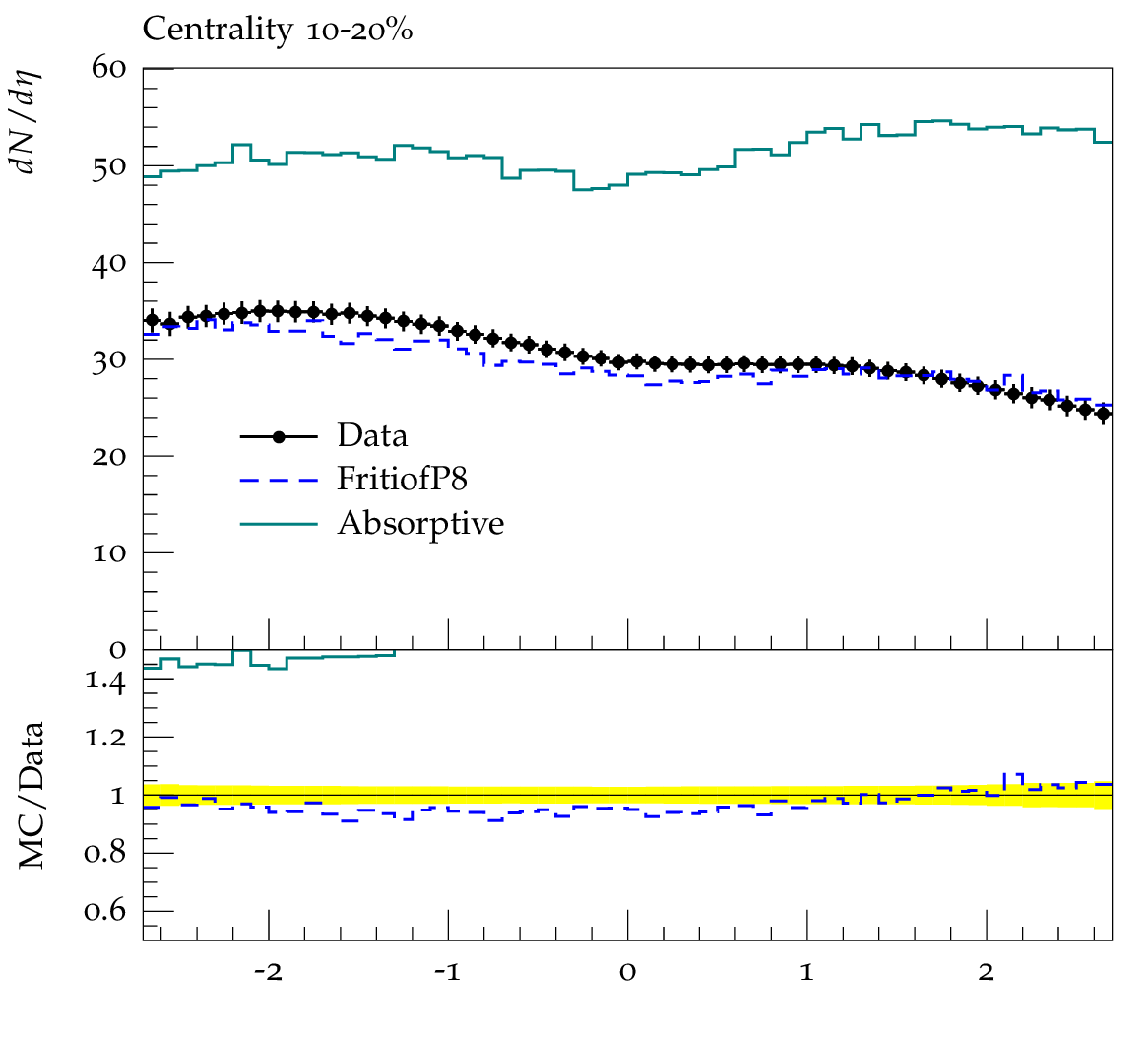}

\begin{minipage}{1.0\linewidth}
  \begin{center}
    (c)
  \end{center}
\end{minipage}
\includegraphics[width=0.5\linewidth,angle=0]{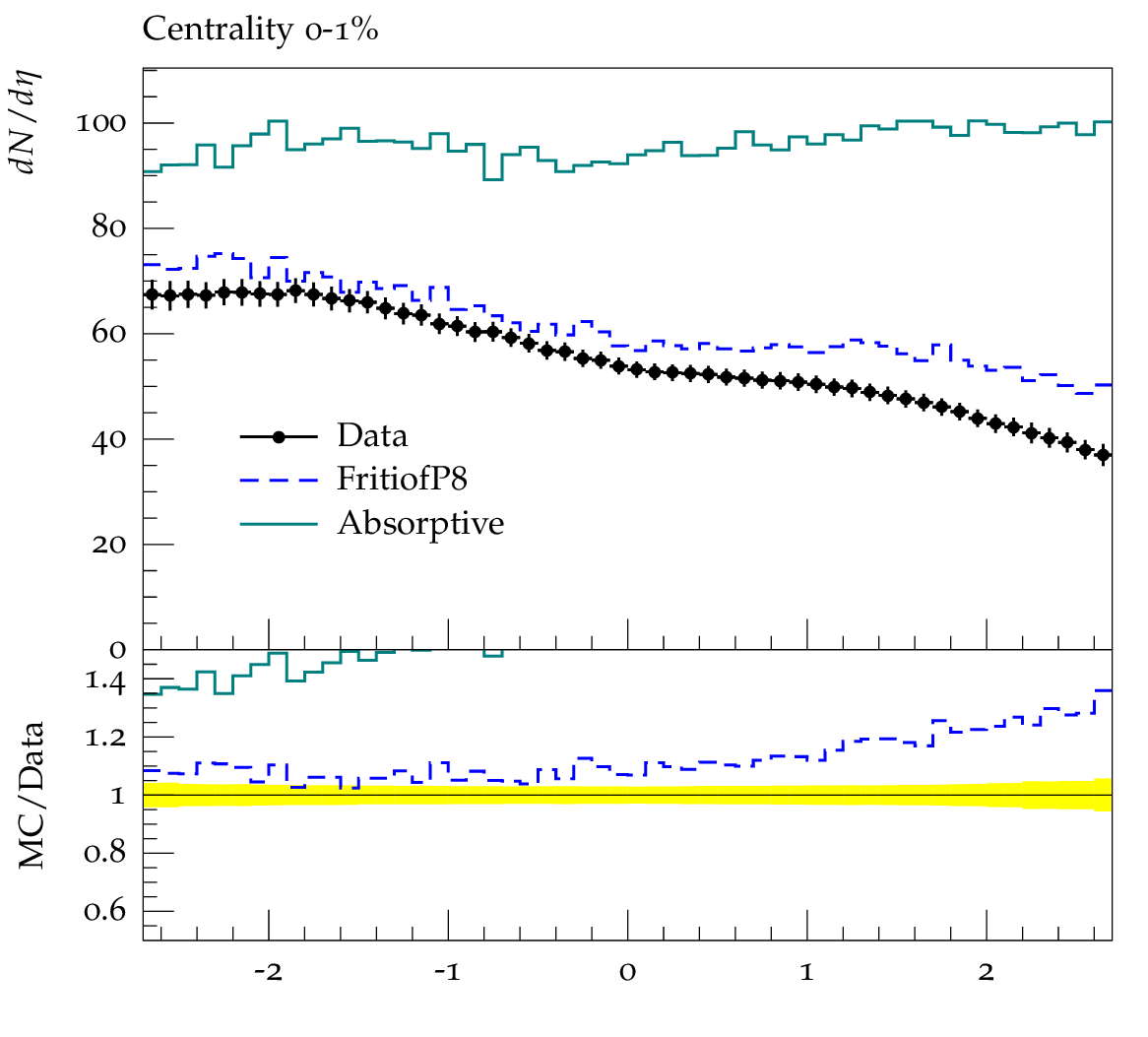}

\caption{Pseudo-rapidity distribution of charged particle multiplicity
  for centralities $60-90\%$ (a), $20-30\%$ (b), and $0-1\%$ (c),
  compared to ``\ourabsorptive'' and ``\newfrit'' particle production
  models.}
\label{fig:multPeriph}
}
We see that while both models describe the peripheral events reasonably well
(which is expected), the new \newfrit model based on diffractive
excitation does a much better job describing both average multiplicity
and the forward--backward asymmetry, as expected.

\subsubsection{Inclusive transverse momentum}

We now compare to centrality-inclusive charged particle $p_\perp$
spectra in different ranges of $\eta$ as measured by \cms
\cite{Khachatryan:2015xaa}. In \figref{fig:pTcent}a we show the
transverse momentum distribution for $-1.0 < \eta < 1.0$. We see that
the \newfrit model performs well at low $p_\perp$, while Absorptive
performs well at high $p_\perp$, as expected. 

\FIGURE[t]{
\centering
\begin{minipage}{1.0\linewidth}
  \begin{center}
    (a)
  \end{center}
\end{minipage}
\includegraphics[width=0.5\linewidth,angle=0]{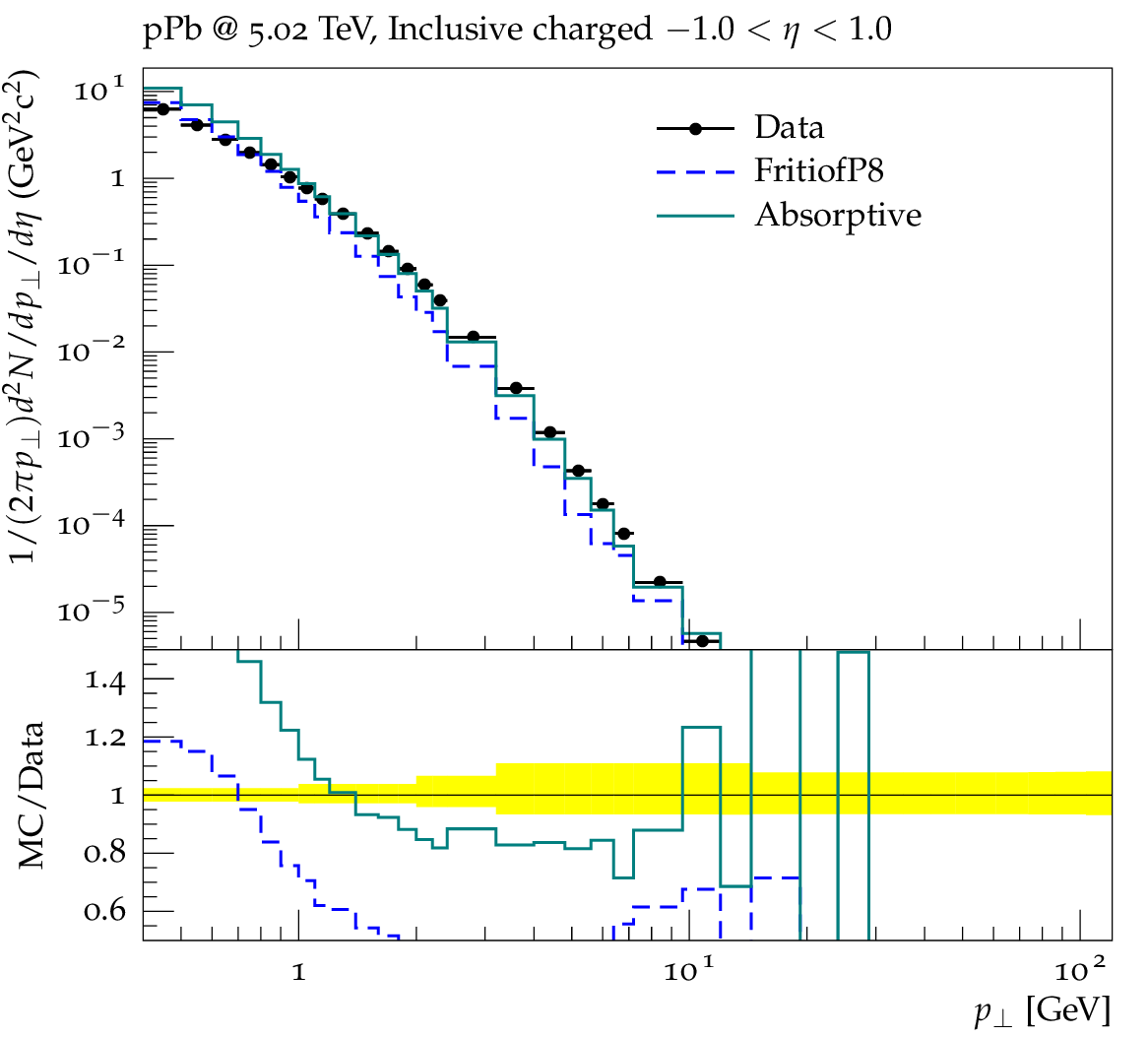}
\begin{minipage}{0.5\linewidth}
  \begin{center}
    (b)
  \end{center}
\end{minipage}\begin{minipage}{0.5\linewidth}
  \begin{center}
    (c)
  \end{center}
\end{minipage}
\includegraphics[width=0.5\linewidth,angle=0]{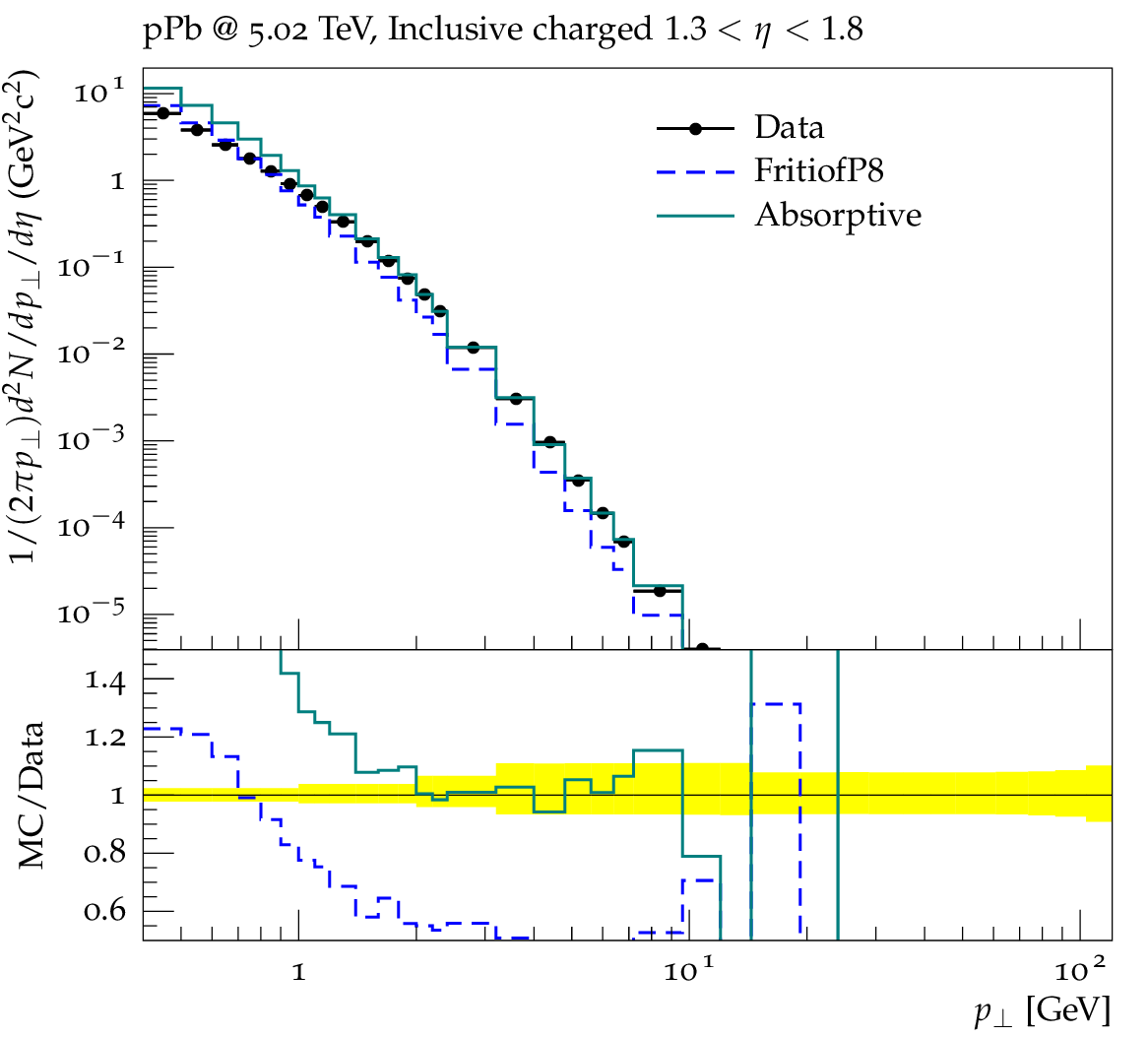}%
\includegraphics[width=0.5\linewidth,angle=0]{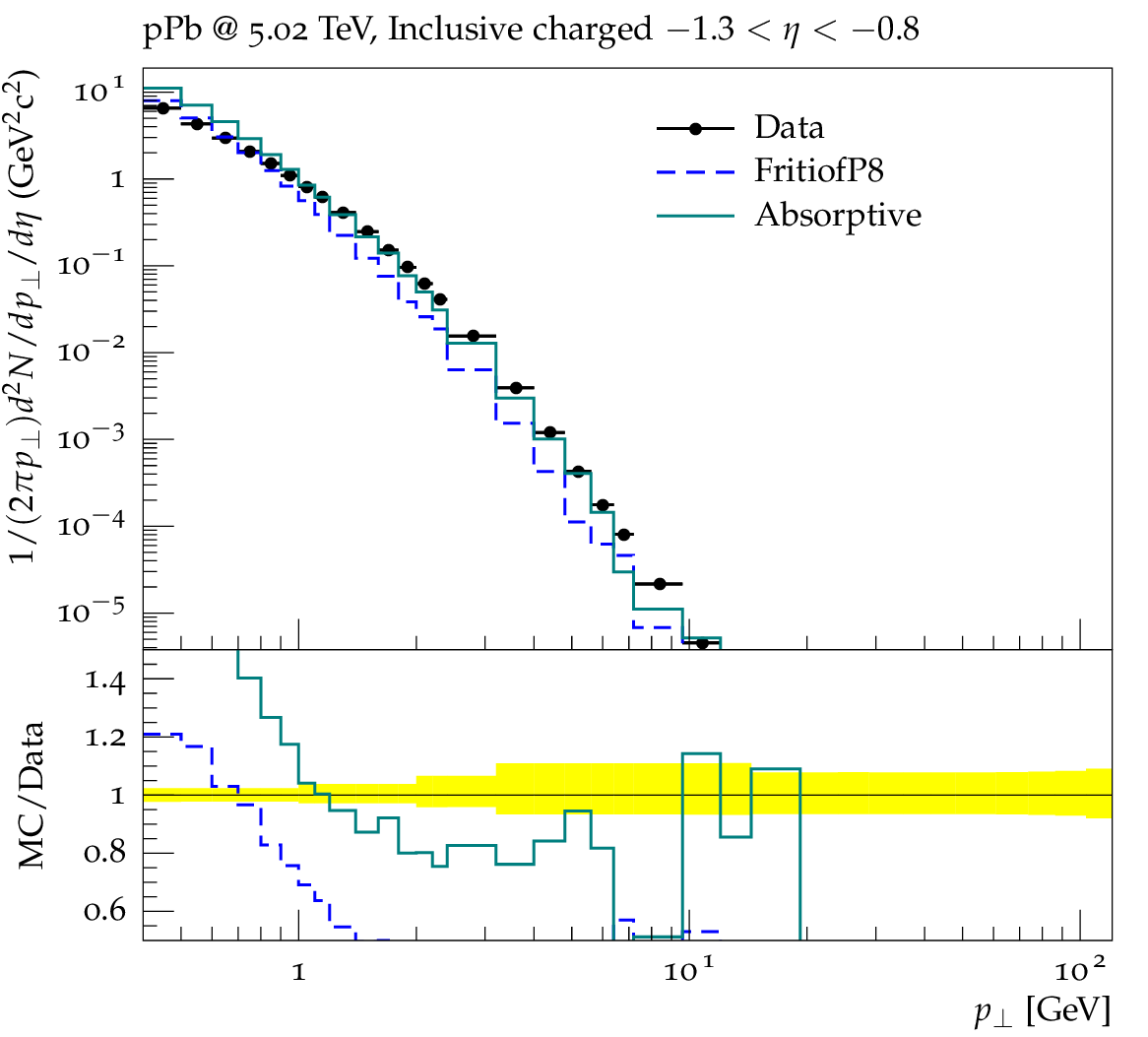}

\caption{Distribution in $p_\perp$, centrality inclusive, for charged
  particles in (a) the central region, $-1.0 < \eta < 1.0$; (b) the
  proton direction, $1.3 < \eta < 1.8$; and (c) the nucleus direction
  $-1.3 < \eta < 0.8$.}
\label{fig:pTcent}
}

The same picture is seen when going
to large negative $\eta$ (\figref{fig:pTcent}c), but performance
of the \newfrit model improves slightly when going to large positive $\eta$
shown in \figref{fig:pTcent}b for $1.3 < \eta < 1.8$ (the
proton side). The "absorptive" model performs as before, but it is
rather surprising that the \newfrit model performs poorly at high
$p_\perp$ here. One explanation could be that the parton distribution
function used for for secondary absorptive events is a Pomeron PDF and
not a proton PDF, due to the fact that secondary absorptive events are
modelled as single diffractive events. This is also a possible
explanation for the poor performance at high $p_\perp$ in
\figref{fig:pTcent}a and \figref{fig:pTcent}c.

\subsection{Uncertainties}
The method presented here for generating final states in \pA\ is
interesting, as it gives qualitatively correct description of the
multiplicity in both the proton and the nucleus direction. It is still
mostly a proof-of-principle since, as we will demonstrate here, using
the \pytppp default settings introduces several hidden assumptions. We
will discuss these assumptions by giving a rough estimate of the
uncertainty associated with each of them. That uncertainty will
decrease, or vanish entirely, when the assumptions are dealt with more
carefully, one by one, which will be done in one or more future
publications.

\subsubsection{PDFs and MPI activity}
\label{sec:pdf-uncert}

In the previous section we described how the secondary absorptive
sub-collisions are approximated as single diffractive excitation
events. The perturbative handling of single diffractive events at high
masses in \pytppp relies on a factorised Pomeron approach, where the
diffractive state is modelled by a Pomeron--proton collision,
including MPI, and we will study two important uncertainties here.
\begin{itemize}
\item The Pomeron PDFs used in the MPI machinery are not really
  appropriate in our model of the secondary absorptive sub-collisions,
  since it is still really the parton density in the proton which
  should drive the MPI. To see possible effects, we have tried to make
  the Pomeron more proton-like, by modifying the PDF used in \pytppp
  to have much more small-$x$ gluons\footnote{The default Pomeron PDF
    in \pytppp is H1 2006 Fit B LO\cite{Aktas:2006hy}, we here use
    instead a simple $Q^2$-independent distribution on the form $xf(x)
    \propto x^a(1-x)^b$, with $a = -0.5$ and $b = 6.0$ for gluons and
    $a = -0.05$ and $b = 0.05$ for quarks.}. This will increase MPI
  activity.
\item Another way of modifying the MPI activity is to change the
  Pomeron--proton cross section used in \pytppp. This is not a
  physical cross section, but rather a free parameter in the program
  which only affects MPI activity, and is adjusted to fit data. The
  default value of this parameter is $10$ mb. We here increase it to
  it's maximal allowed value, $40$ mb, to better reflect a $pp$
  absorptive cross section.
\end{itemize}

In \figref{fig:pdfMult} we show the pseudo-rapidity distribution of
charged particles for the highest centrality bin. The two variations
above are labelled ``PDF'' and ``MPI'' respectively while \newfrit is
the same as before.

\FIGURE[t]{
\centering
\includegraphics[width=0.6\linewidth,angle=0]{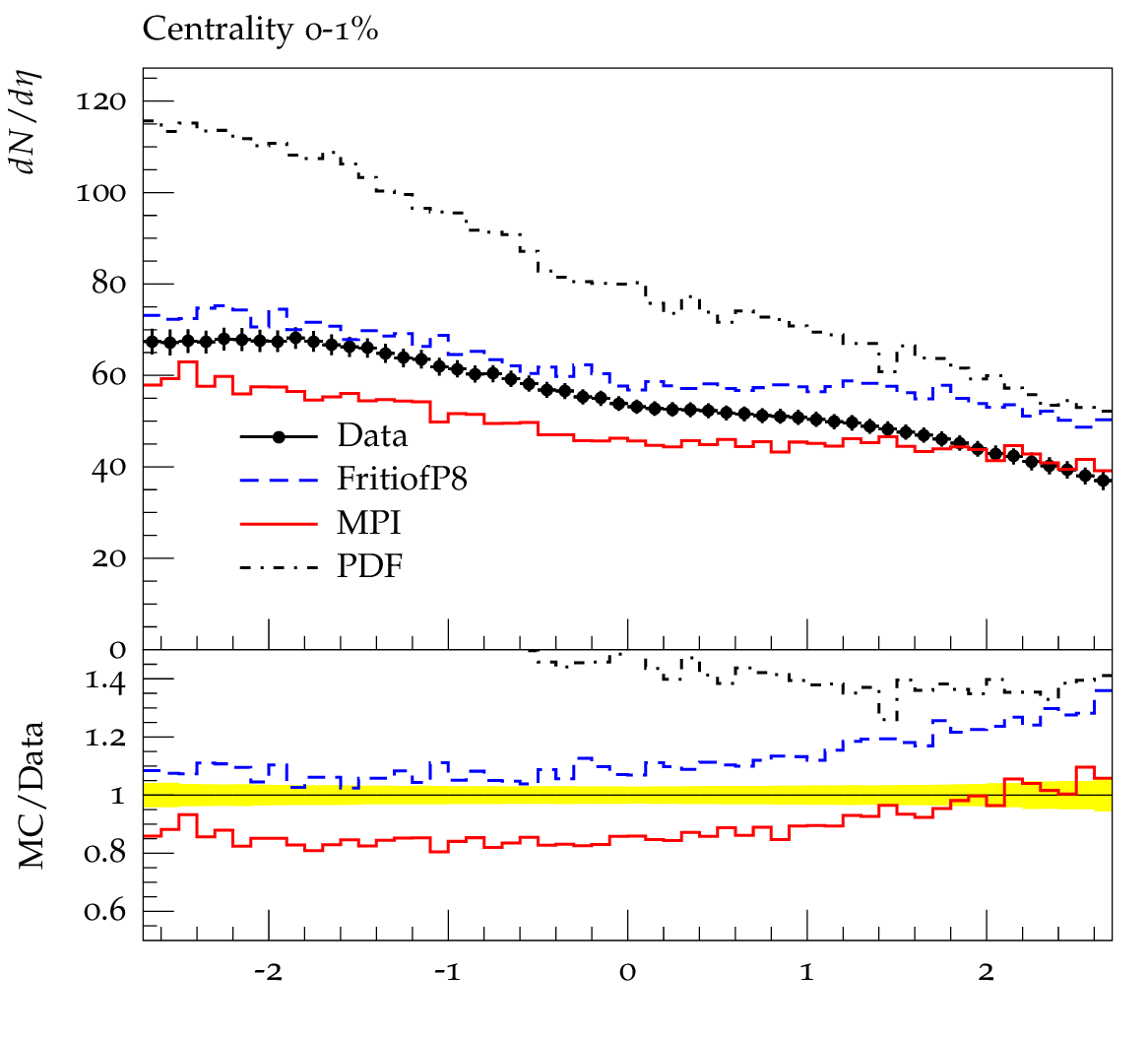}
\caption{Pseudo-rapidity distribution of charged particle multiplicity
  for centrality $0-1\%$ compared to three different ways of
  estimating the number of MPIs, giving an estimate of the method
  uncertainty.}
\label{fig:pdfMult}
}

In total, the envelope of the three lines gives what we believe to be
a reasonable, albeit conservative, estimate of the uncertainty so far
associated with the approximations regarding the parton densities and
the amount of MPIs.

\subsubsection{GG uncertainty}
In \sectref{sec:comp-with-dipsy} we described how the Glauber--Gribov
cross section fluctuations could be parameterised with a log-normal
distribution. This new parameterisation was used in all the previous
data comparison plots, here we show how the traditional
parameterisation (also fitted to \pp~data) compared to the new
one. Since distributions of wounded nucleons for the two models
differs most in the tail, we are most sensitive in central collisions,
In \figref{fig:gccfMult} we show the uncertainty in central particle
production arising from changing the parametrisation of cross section
fluctuations. We see that the uncertainty covers data well, but is
smaller than the uncertainties in the handling of secondary absorptive
events above.

\FIGURE[t]{
\centering
\includegraphics[width=0.6\linewidth,angle=0]{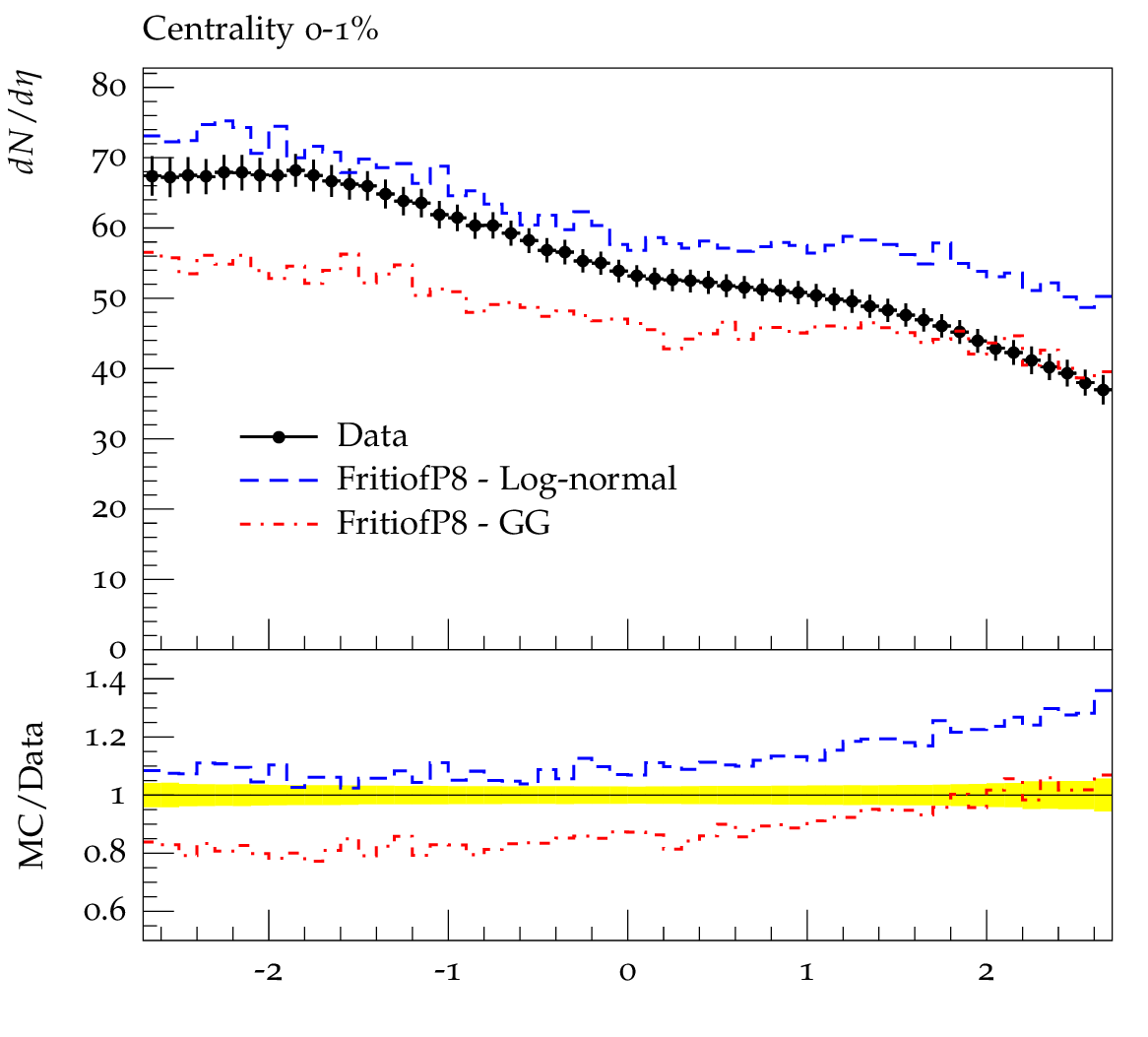}
\caption{\label{fig:gccfMult} Pseudo-rapidity distribution of charged
  particle multiplicity for centrality $0-1\%$ compared to two
  parameterisations for the GG model.}
}

\section{Conclusions and outlook}
\label{sec:outlook}

Collisions of heavy nuclei with protons or each other, are often
understood in terms of their deviation from minimum bias \pp\
collisions, which after many years of work on models for multiple
partonic interactions, are fairly well understood. The extrapolation
from \pp\ to \pA\ or \AA, which is usually based on Glauber models, is
therefore a crucial step toward understanding the deviations, and thus
crucial for understanding how and if a de-confined plasma of quarks
and gluons are created in heavy ion collisions, and what kind of
impact it has on final state observables. In this article we have
discussed Glauber models, their extensions including fluctuations and
the impact these extensions on the number of wounded nucleons,
followed by a possible way of using that knowledge to generate full,
exclusive final states as a sum of \pp\ collisions. We will conclude
on these two parts separately, followed by an outlook primarily
focused on extending the model for full final states, to also include
microscopic models for collective effects.

\subsection{Fluctuations in the Glauber formalism}

Following Good and Walker, we have discussed diffractive excitation as
a manifestation of fluctuations in the substructure of the nucleons.
As a result of this discussion we identified the \textit{inclusively
  wounded cross section} to be the relevant \pp\ cross section for
calculations of the number of wounded nucleons. The wounded cross
section has contributions from absorptive processes plus diffractively
excited target nucleons, in brief form we have for the semi-inclusive
\pp\ cross sections:

\begin{equation}
  \sigma\subwinc=\sigma\subabs + \sigma\subdd + \sigma\subsdt = 
  \sigma\subtot - \sigma\subel - \sigma\subsdp. 
\end{equation}

We have discussed the developments in calculating the number
distribution of inclusively wounded nucleons in the so-called
Glauber--Gribov approach by Strikman and co-workers. By comparing
distributions of the cross section to ones calculated with the \dipsy
model and measurements, we find that the parametrisation of the cross
section suggested in the Glauber--Gribov approach does not fully
include all fluctuations necessary to describe the ones in \dipsy.

The simplest Glauber calculations using a fixed black disk, can
clearly not describe any fluctuations, and we want to emphasize that
just setting $\sigma_{\NN} = \sigma\subwinc$ in such a calculation, is
not enough if one wants to calculate the contribution from the wounded
nucleons to a centrality defining observable.

We have shown that a Glauber--Gribov calculation with a black disk
fluctuating in size, can be used to predict the distribution of
inclusively wounded nucleons, if the fluctuating black disk is
attributed to fluctuations in the projectile, while averaging over
fluctuations in the target nucleon. However, if one wants to separate
absorptively wounded nucleons from diffractively wounded ones, it is
necessary to also consider fluctuations in the individual target
nucleons.

We have suggested a new functional form for the fluctuations in the
\pp~interaction, with a higher tail to a larger cross section.
Instead of the parametrization introduced by Strikman \textit{et al.},
and used in several experimental analyses, we suggest a log-normal
distribution, which is believed to give a somewhat better description
of the inclusive distribution, though not necessarily a more realistic
picture of the cross section fluctuations.  We have also included
fluctuations in the target nucleons by introducing a crude
Glauber-like model, where the radii of the projectile and target are
allowed to fluctuate independently between two values. The model
includes four parameters, and can be fitted to describe four
independent semi-inclusive \pp\ cross sections, including the
inclusive wounded one. By using this model to include projectile
fluctuations in the Glauber--Gribov approach, we separate the
inclusively wounded nucleon into absorptively and diffractively
wounded ones.

The parametrisations and toy-models studied are built on the
assumption that distributions of wounded nucleons can be described
solely by fluctuations in the proton size; projectile size for the
inclusive distributions, and target nucleon size for distinguishing
between absorptively and diffractively wounded nucleons.  Thinking
about the dynamics of more involved calculations, like the \dipsy
model used in this paper, it is clear that size fluctuations cannot
account for all the relevant physics.  In the \dipsy model,
fluctuations in cross section will also arise when \eg~a small
projectile is very dense, and therefore gives rise to a larger cross
section. Similarly, a large projectile can be dilute, giving rise to a
smaller cross section.  In the language of this article, such effects
would need to be accounted for by a fluctuating, $b$-dependent
opacity, resulting in a profile function which would go beyond the
simple Gaussian, grey -or black disks, but still not be as
calculationally involved as the full, dynamical models. This will be
the subject of a future publication.

\subsection{Full final states}

We have given a proof-of-principle for an approach to model exclusive
final states in \pA\ collisions as a sum of several \pp\
collisions. The approach uses \pytppp to calculate the hardest
absorptive sub-collision as a normal non-diffractive \pp\ collision,
while the subsequent absorptive collisions are modelled as single
diffractive events. This was inspired by the old Fritiof model, which
is valid at lower energies, but adds another dimension, as high-mass
single diffractive exchanges can now be treated perturbatively and
allows for multiple parton--parton scatterings.

We have shown that this approach, in a quite crude implementation, is
able to give a reasonable description of some recently published
final-state measurements of \pPb\ collisions at the LHC, but the
uncertainties in our approach are quite large. In future studies we
will try to eliminate these uncertainties, in the hope to get a
theoretically well motivated and more accurate description of data.

To do this it is helpful to have data published in a usable form for
comparison with event generators. The LHC \pp\ community has come a
long way in this respect by publishing many of their results in the
form of Rivet routines. In this way the measurements are presented in
a clearly reproducible form, including all relevant kinematical cuts
and unfolding of detector effects but free from dependence of
theoretical models. In order to allow for the development of better
event generators for heavy ion collisions, it is imperative that the
experimental heavy ion community adapts a similar way of presenting
their results.

\subsection{Outlook: modelling collective effects}

In both the theoretical and experimental communities around heavy ion
collisions, much attention is given to observables thought to convey
information about a possible plasma state created in the collision. A
direct application of the work presented in this article, is to use
the final state extrapolation as a baseline for calculations of
collective effects using microscopic, QCD based models. In the \pp\
community, much attention have been given to models of final state
interactions including colour reconnections, rope hadronisation and
junction formation
\cite{Bierlich:2014xba,Bierlich:2015rha,Christiansen:2015yqa}. These
models have, in \pp\, been shown to reproduce: the enhancement of
strange hadrons to non-strange hadrons in dense environments
\cite{Adam:2016emw}; flow-like effects in hadron ratios as function of
$p_\perp$; and preliminary studies have shown that interactions
between strings can produce a ridge \cite{Altsybeev:2015vma}.

To implement such models in a framework like the one presented here, a
necessary component is a good understanding of the sub-collisions in
transverse space. Such a picture is not included in \eg\ the \pytppp
MPI model, and must therefore be added \textit{a posteriori}. Guidance
can then be had from the \dipsy model, which includes detailed
information about the transverse space structure, but does not produce
final states describing data as well as \pytppp.

Finally, the final state model introduced here should be developed
further to model fully exclusive hadronic final states also in \AA\
collisions. Since every projectile here also becomes a target, we
suspect that the model cannot be transferred one-to-one, but that some
modifications may be needed. Here modelling collective effects using
microscopic models is also highly desirable.

\section*{Acknowledgments}

This work was supported in part by the MCnetITN FP7 Marie Curie
Initial Training Network, contract PITN-GA-2012-315877, the Swedish
Research Council (contracts 621-2012-2283 and 621-2013-4287).

\bibliographystyle{utcaps}
\bibliography{refs}

\end{document}